\documentclass[journal]{IEEEtran}
\usepackage{graphicx}
\usepackage{cite}
\usepackage{picinpar}
\usepackage{amsmath}
\usepackage{url}
\usepackage{flushend}
\usepackage[latin1]{inputenc}
\usepackage{colortbl}
\usepackage{soul}
\usepackage{multirow}
\usepackage{pifont}
\usepackage{color}
\usepackage{alltt}
\usepackage[hidelinks]{hyperref}
\usepackage{enumerate}
\usepackage{siunitx}
\usepackage{breakurl}
\usepackage{epstopdf}
\usepackage{pbox}
\usepackage{makecell}
\usepackage{amssymb}
\ifCLASSINFOpdf
\else
\fi
\begin{document}
%
\title{Decision Making of Connected Automated Vehicles at An Unsignalized Roundabout Considering Personalized Driving Behaviours}
%
%
%

\author{Peng Hang,
        Chao Huang, Zhongxu Hu, Yang Xing, and Chen Lv
\thanks{This work was supported by the SUG-NAP Grant (No. M4082268.050) of Nanyang Technological University, Singapore.}
\thanks{P. Hang, C. Huang, Z. Hu, Y. Xing and C. Lv are with the School of Mechanical and Aerospace Engineering, Nanyang Technological University, Singapore 639798. (e-mails: \{peng.hang, chao.huang, zhongxu.hu, xing.yang, lyuchen\}@ntu.edu.sg). Corresponding author: C. Lv}
\thanks{This work has been submitted to the IEEE for possible publication. Copyright may be transferred without notice, after which this version may no longer be accessible.}
}

%
%

\markboth{ }
{Shell \MakeLowercase{\textit{et al.}}: }
%



\maketitle

\begin{abstract}
To improve the safety and efficiency of the intelligent transportation system, particularly in complex urban scenarios, in this paper a game theoretic decision-making framework is designed for connected automated vehicles (CAVs) at unsignalized roundabouts considering their personalized driving behaviours. Within the decision-making framework, a motion prediction module is designed and optimized using model predictive control (MPC) to enhance the  effectiveness and accuracy  of the decision-making algorithm. Besides, the payoff function of decision making is defined with the consideration of vehicle safety, ride comfort and travel efficiency. Additionally, the constraints of the decision-making problem are constructed. Based on the established decision-making model, Stackelberg game and grand coalition game approaches are adopted to address the decision making of CAVs at an unsignalized roundabout. Three testing cases considering personalized driving behaviours are carried out to verify the performance of the developed decision-making algorithms. The testing results show that the proposed game theoretic decision-making framework is able to make safe and reasonable decisions for CAVs in the complex urban scenarios, validating its feasibility and effectiveness. Stackelberg game approach shows its advantage in guaranteeing personalized driving objectives of individuals, while the grand coalition game approach is advantageous regarding the efficiency improvement of the transportation system.
\end{abstract}


\IEEEpeerreviewmaketitle

\section{Introduction}
\subsection{Motivation}
\IEEEPARstart{T}{o} further advance the safety and efficiency of future mobility systems, vehicle coordination and management in complex conditions, particularly in urban scenarios, are worthwhile investigating. For urban scenarios, unsignalized roundabout intersections are usually considered to be more complex and challenging than crossroad intersections, with respect to multi-vehicle interactions [1]-[3]. A roundabout is defined as a circular intersection in which all vehicles travel around a circular island at the center with the counter-clockwise direction (driving on the right) [4]. Comparing to crossroad intersections, traffic lights are not necessary for roundabouts to control the traffic flow. Therefore, for the entering vehicles, a complete stop may not be required. As a result, the traffic delay can be  decreased, and the traffic capacity can be improved [5]. For human-driven vehicles, the traffic rules at roundabouts are defined as follows: (1) Circulating vehicles have the priority over entering and merging vehicles; (2) Exiting vehicles have the priority over entering vehicles; (3) Large vehicles have the priority over small vehicles. The aforementioned rules are able to bring remarkable advances to the traffic efficiency and safety at roundabouts [6]. Even so, however, with increasing traffic flow, especially during peak hours, congestion and conflicts are inevitable due to traffic dynamics, various travel objectives, and individual driving styles of human drivers. In general, decision making and coordination of vehicles at unsignalized roundabouts is still an open challenge for human drivers as well as traffic management.
Connected and automated driving technology has great potential to deal with the issues of traffic congestion and conflicts and ensure driving safety in complex urban scenarios [7]. In a connected driving environment, the motion states, surrounding traffic situations, and even the intentions and planned behaviors of driving can be shared between connected automated vehicles (CAVs). This enables the correct, reasonable and effective decisions to be made for multi-vehicle interactions, particularly in complex situations [8]. As a result, the driving safety, travel efficiency and riding comfort of CAVs can be improved.

\subsection{Related Works}
To deal with decision making of CAVs at intersections, the concept of centralized traffic management system, in which a centralized controller is proposed to manage CAVs within the controlled area, has been widely studied [9]-[11]. In such cases, when a vehicle enters the controlled area, it must hand over the control authority to the centralized controller. Then, the centralized controller guides the vehicle to pass through the intersection according to an optimized passing sequence [12]. In [13], a distributed coordination algorithm is developed for dynamic speed optimization of CAVs in the urban street networks for improving the efficiency of network operations. However, the turning behavior of CAV is not considered in the traffic networks. In [14], a distributed conflict-free cooperation approach is proposed to optimize the traffic sequence of multiple CAVs at the unsignalized intersection for managing CAV operations with a high efficiency, but it also requires a large amount of computation. To make a good trade-off between the system performance and computation complexity, a tree representation approach, which is combined with the Monte Carlo tree search and some heuristic rules, is proposed to find the global-optimal passing sequence of CAVs [15].

The aforementioned studies mainly focus on global optimization of the passing sequence and velocity coordination for CAVs at intersections. Although the centralized management system can improve the traffic efficiency, the computation burden largely increases with the rise of the amount of vehicles, which brings challenges in reliability and robustness [16]. However, if there is no global coordination at the intersection, then it is also a challenge to individual vehicles during their interactions with others. Therefore, the individual decision making ability is quite important for CAVs [17]. To address the dynamic decision making issue of CAVs at a roundabout intersection, a Swarm Intelligence (SI)-based algorithm is designed. In this algorithm, each CAV is regarded as an artificial ant that can self-calculate to make reasonable decisions with internet of things (IoT) technique [18]. To improve the reliability, robustness, safety and efficiency of CAVs at intersections, a digital map is used to predict the paths of surrounding vehicles, and afterwards, the potential threats assessed are provided to the ego vehicle to make safe and efficient decisions [19]. In [20], a cooperative yielding maneuver planner is designed, which allows CAVs circulating inside single-lane roundabouts to create feasible merging gaps for oncoming vehicles. In [21], Bayesian Gaussian mixture model is combined with nonlinear model predictive control to realized decision making and collision avoidance for CAVs. In addition, reinforcement learning (RL) is an effective approach to address the decision-making issue of CAVs at intersections. In [22], several state-of-the-art model-free deep RL algorithms are implemented into the decision-making framework, which advances the robustness when dealing with complex urban scenarios. In [23], an adaptive RL based decision-making algorithm is studied, and an underlying optimization-based trajectory generation module is designed to improve the effectiveness of the decision making. In [24], a novel optimization algorithm is investigated for multi-agent decision making, where the future position of moving obstacles is predicted, and thus potential collisions are avoided effectively. However, learning-based approaches are data-driven ones, thus their performances are affected by the quality of the dataset.

In addition to the above approaches, game theoretic approach shows superiority and effectiveness in interaction modeling and decision-making of CAVs [25]. In [26], a decision making algorithm is designed for CAV control at roundabout intersections with game theoretic approach. However, only two players, i.e., the ego vehicle and an opponent vehicle, are considered. In [27], the level-k game theory is applied to model multi-vehicle interactions at an unsignalized intersection, and this approach is combined with the receding-horizon optimization and imitation learning to design the decision-making framework for CAVs. In general, the game theoretic approaches can be applied not only to handle the decision making of CAVs, but also to simulate the interactive behaviors of intelligent agents [28].

In existing studies of decision making for CAVs, personalized driving is seldom considered. Actually, different passengers have distinguished demands for their travel, which can be reflected by different driving styles of CAVs [29]. In [30], a neural network approach is used to establish the personalized driver model, and the aggressiveness index is proposed to evaluate the driving style. In [31], the driving style is divided into three types, including aggressive, moderate and conservative types, based on energy consumption analysis. Considering the human-like driving features, a risk-based autonomous driving motion controller is designed, which can guarantee both the safety and comfort [32]. In this paper, with consideration of personalized driving styles, a game theoretic decision-making framework is designed to address the decision-making issue of CAVs at an unsignalized roundabout.

\subsection{Contribution}
To further improve the safety and efficiency of the intelligent transportation system while considering personalized driving behaviours of individual users, two game theoretic approaches are developed to address the decision making problem of CAVs in complex urban scenarios. The contributions of this paper are summarized as follows: (1) A game theoretic decision-making framework is proposed to deal with the merging, passing and exiting of CAVs at an unsignalized roundabout zone. Both the individual benefit of single vehicle and the social benefit of the entire transportation system are considered. The stackelberg game approach is in favor of personalized driving of individuals, while the grand coalition game approach is advantageous with respect to the efficiency improvement of the entire system; (2) The motion prediction of CAVs is considered in the decision-making framework using MPC to enhance the effectiveness of the decision making; (3) The interactions and decision-making of CAVs with different driving styles are studied, which provides personalized driving experience for passengers in terms of driving safety, ride comfort and travel efficiency.
\subsection{Paper Organization}
The remainder of the paper is organized as follows. The problem formulation and system framework of the decision making for CAVs at an unsignalized roundabout are introduced in Section II. In Section III, the motion prediction model of CAVs is constructed. In Section IV, the game theoretic decision-making algorithm is studied to address the decision-making of CAVs. The testing results and analysis are presented in Section V. Finally, Section VI concludes the work of this paper.

\section{Problem Formulation and System Framework}
\subsection{Decision Making of CAVs at An Unsignalized Roundabout}
As mentioned in Section I, a wide range of studies have been conducted to address the decision-making issue of CAVs at unsignalized roundabouts. However, many of them limit the driving scenario to a single-lane roundabout, which is only able to deal with the decision making between two vehicles which are in the main lane and the round lane, respectively. The overtaking, lane change, merging, and other interaction behaviors among multi vehicles at unsignalized roundabouts have rarely been reported. In this paper, to further advance the algorithm and increase the system complexity, a two-lane unsignalized roundabout scenario with multi-vehicle interactions is investigated.

Moreover, different users would have distinguished demands during their trips in terms of vehicle safety, ride comfort and travel efficiency. And this can be reflected by different driving styles of CAVs. Therefore, for future CAV design, it is reasonable to consider personalized preferences and human-like driving behaviours. Currently, most of the related studies have not yet considered driving styles and their effects on CAVs decision making. In this paper, different driving styles are embedded in the modelling phase of CAVs, which introduces additional dimension and features to the decision making module of CAVs.

The designed scenario of an unsignalized roundabout is illustrated in Fig. 1. In this scenario, all vehicles are assumed to be CAVs in a connected driving environment. The unsignalized roundabout consists of eight two-lane main roads and one two-lane round road, i.e., $M1$, $M2$, $M3$, $M4$, $\hat{M}1$, $\hat{M}2$, $\hat{M}3$, $\hat{M}4$ and $RR$. The eight main roads are linked with the round road via four entrances and four exits, which are denoted by $A_{in}$, $B_{in}$, $C_{in}$, $D_{in}$, $A_{out}$, $B_{out}$, $C_{out}$ and $D_{out}$. In Fig. 1, the red car is the host vehicle (HV); the blue cars are the neighbor vehicles (NVs) of HV; the purple cars are the lead vehicles (LVs), and the black cars are the irrelevant vehicles (IVs). The effect of LV on HV is unidirectional. Any decision-making result of HV will not lead to the change of LV. While, NVs are opponents to HV, and the decision-making behaviors are interactive. Considering that there is no direct interaction between IV and HV, we have the following assumption that the effects of IVs' driving behaviors on HV's decision making are neglected in this work. It should be noted that the name of NV, LV and IV are not fixed. With the change of HV's relative position, the roles of surrounding vehicles would change.

\begin{figure}[t]\centering
	\includegraphics[width=8.5cm]{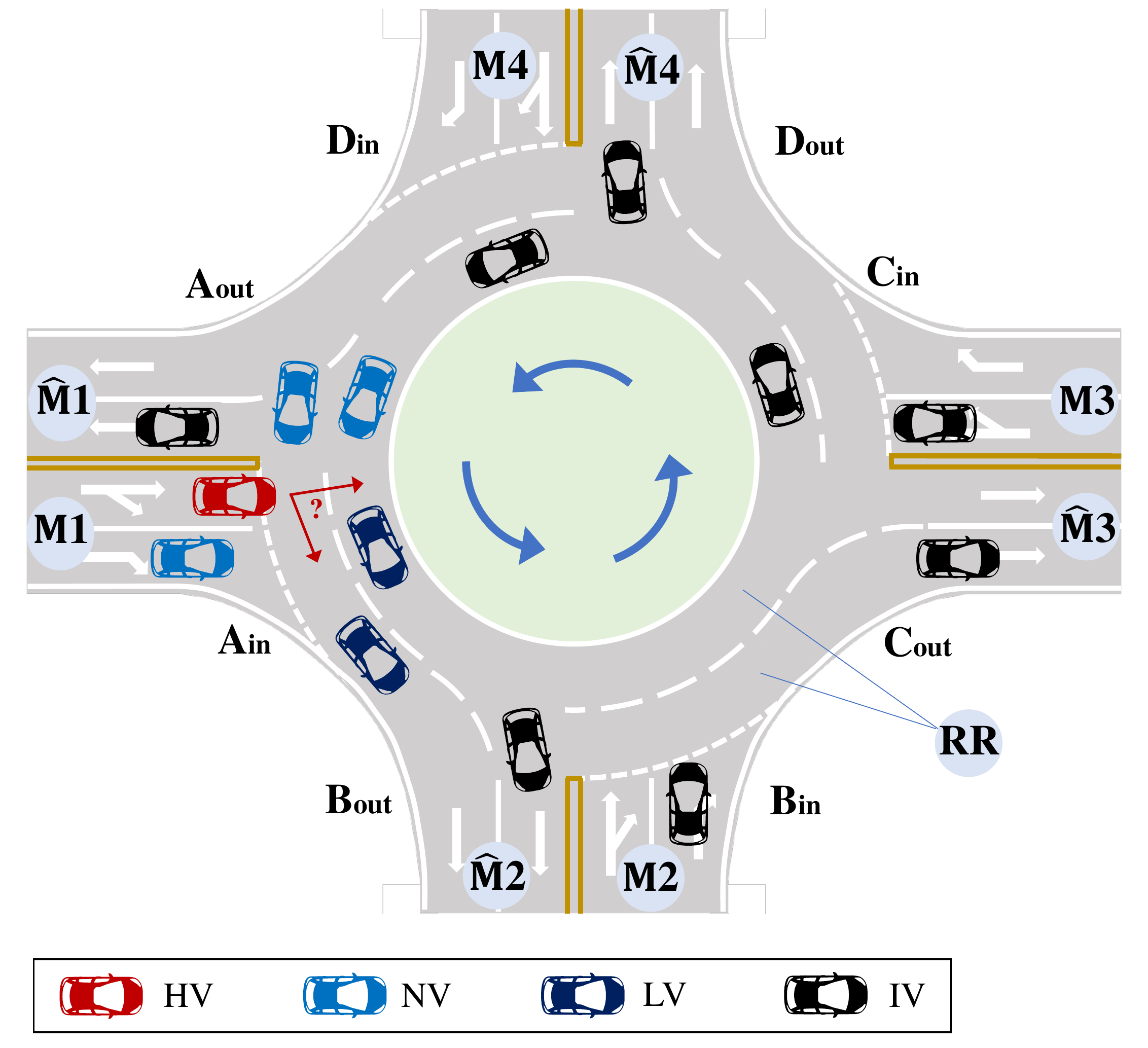}
	\caption{The scenario of the unsignalized roundabout with CAVs.}\label{FIG_1}
\end{figure}

The decision making process of CAVs at the roundabout can be classified into three stages, i.e., the entering, passing and exiting, which are illustrated in Fig. 2. Fig. 2 (a) presents the entering scenario, in which HV on the main road $M1$ wants to enter the round road $RR$ from the entrance $A_{in}$. It needs to interact with its surrounding vehicles and make a decision on its merging behaviour. For HV, three choices are presented, i.e., merging into the outside lane of round road $RR$ , merging into the inside lane of round road $RR$, lane keeping until stopping. The decision making of HV is remarkably influenced by the motion states and driving behaviors of surrounding vehicles, i.e., NV1, NV2, NV3, LV1, and LV2. For instance, if HV decides to merge into the outside lane of round road $RR$, it will pursue the right of way with NV1 and NV2. For NV1 and NV2, they can slow down and give ways to HV, or speed up and fight for the right of way. If the driving styles of NV1 and NV2 are conservative, they will choose the former. However, if NV1 and NV2 are the aggressive driving style, they are likely to choose the latter. The decision making results of the three vehicles are associated with many complex factors including the motion states, relative positions and especially the driving styles.

\begin{figure}[t]\centering
	\includegraphics[width=8.5cm]{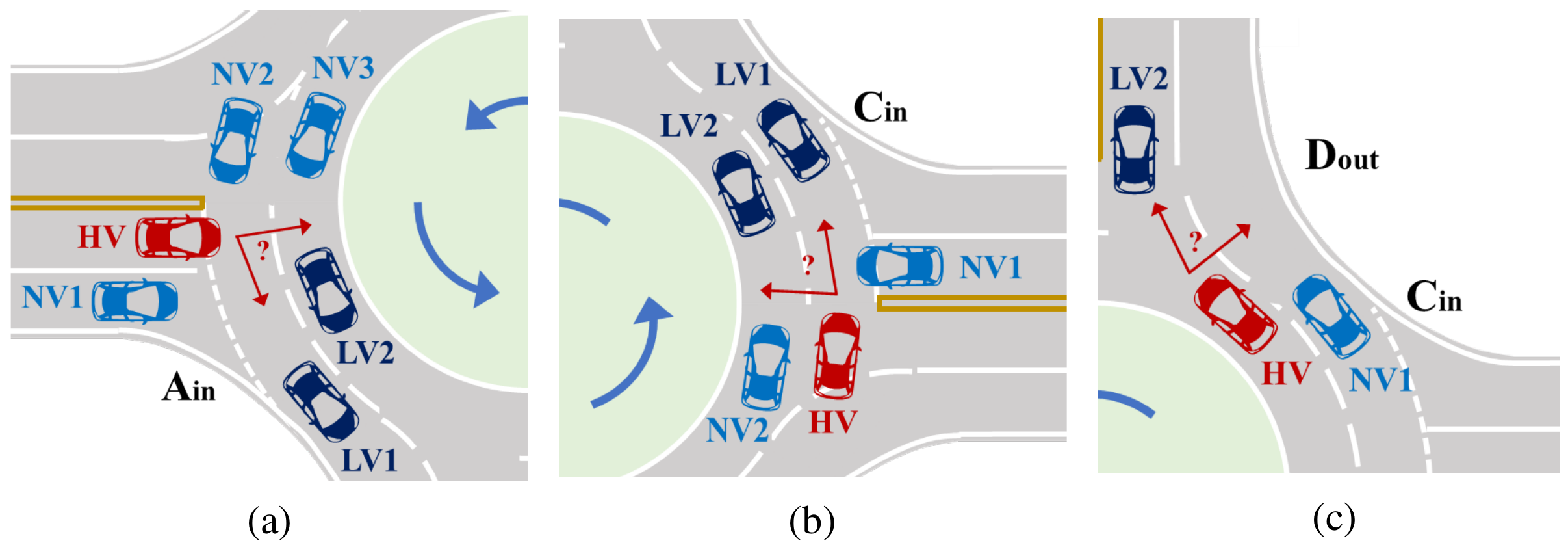}
	\caption{Decision making process of CAVs at the roundabout: (a) Entering; (b) Passing; (c) Exiting.}\label{FIG_2}
\end{figure}

After merging into the round road, it comes to the passing stage, in which HV can change lanes due to the low speed of LV. Meanwhile, it has to deal with the merging and overtaking of other vehicle. As Fig. 2 (b) shows, NV1 on the main road $M3$ wants to enter the round road $RR$ from the entrance $C_{in}$. HV has to make decisions from the following three choices: (1) keeping the lane, slowing down and giving way to NV1, (2) keeping the lane, speeding up and pursuing the right of way with NV1, (3) changing lanes to the inside lane of round road $RR$. If HV chooses the third one, it must interact and game with NV2 besides HV. Fig. 2 (c) shows the exiting stage of HV at the roundabout. In this scenario, HV is assumed to exit from $D_{out}$ and merge into the main road $\hat{M}4$. It has to make the decision that merging into which lane of $\hat{M}4$. Additionally, it must take into consideration the overtaking or cutting in behaviors of surrounding vehicles. In general, the decision making and driving behaviors of CAVs at the unsignalized roundabout are remarkably related to their different driving styles and objectives. This paper aims to study the interactions and decision making strategies of CAVs at the unsignalized roundabout considering personalized driving styles.

\subsection{Decision Making Framework for CAVs}
To address the decision making issue of CAVs at the unsignalized roundabout, a game-theoretic decision making framework is proposed, which is illustrated in Fig. 3. As mentioned above, the decision making results of CAVs are remarkably influenced by the driving style. In view of this, three different driving styles, i.e., aggressive, conservative and normal, are defined for HV and surrounding vehicles [33,34].

(a) Aggressive: This driving style gives the highest priority to the travel efficiency. Drivers would like to take some aggressive behaviors to reach their driving objectives including sudden accelerations or decelerations, frequent lane change or overtaking, and forcibly merging.

(b) Conservative: It means cautious driving. Drivers care more about safety. Therefore, lane keeping, large gap and low speed driving are preferred.

(c) Normal: Most drivers belong to this kind of driving style, which is positioned between the aforementioned two categories, expecting a balance among different driving objectives and performances.

\begin{figure}[h]\centering
	\includegraphics[width=8.5cm]{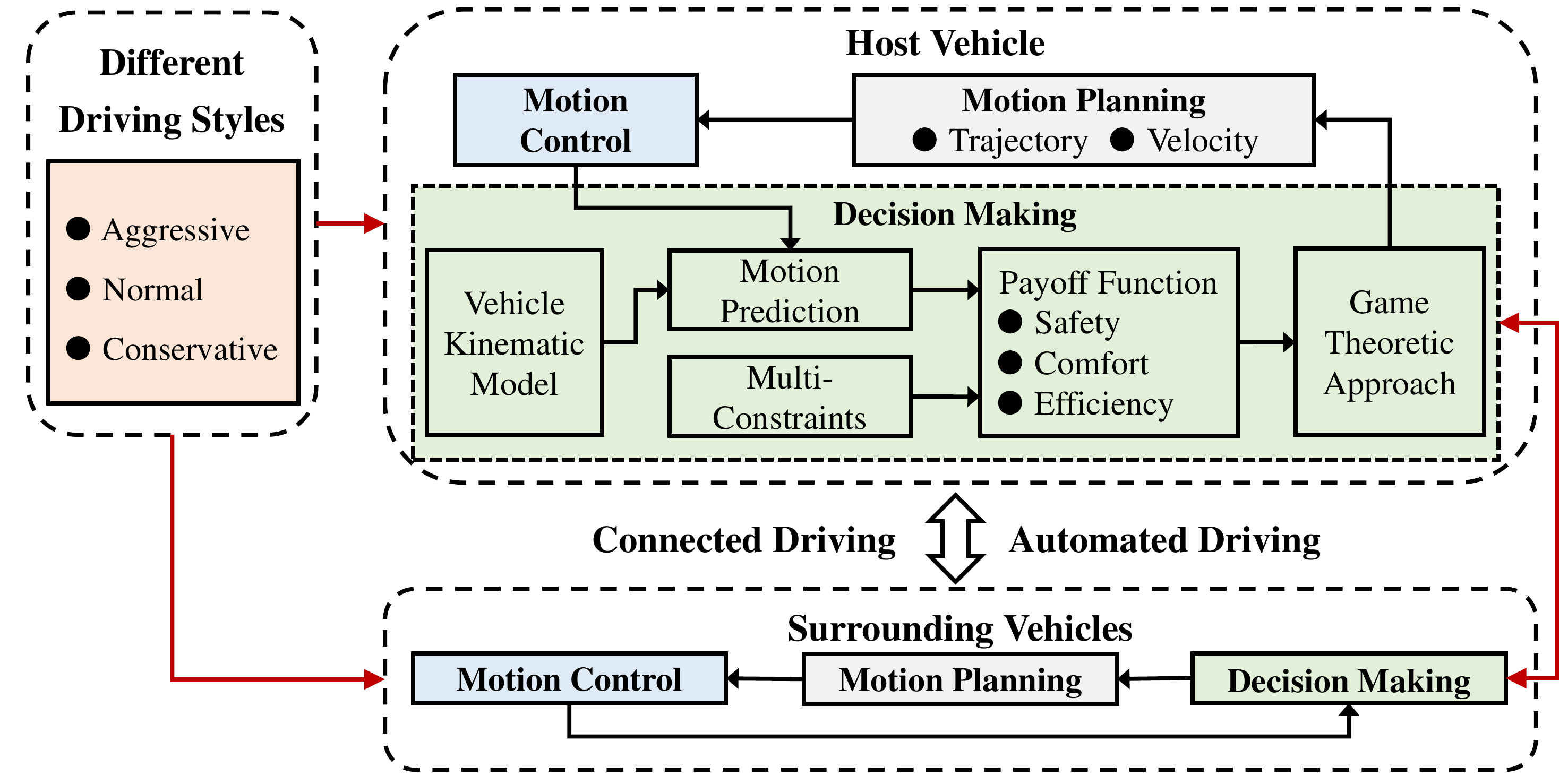}
	\caption{Decision making framework for CAVs.}\label{FIG_3}
\end{figure}

In the decision making algorithm, the vehicle kinematic model is applied to the motion prediction. According to the predicted motion states, the payoff function for decision making is then formulated with multi-constraints, in which the typical driving performances, i.e., safety, comfort and efficiency, are considered. Based on the payoff functions of decision-making for CAVs, the game theoretic approach is utilized to address the interaction and decision making issue of CAVs at the roundabout. After solving the formulated optimization problem of the game, the decision-making results are sent to the motion planning module. Finally, the motion control module conducts the driving behaviors generated from the decision making. In this paper, we mainly focus on the algorithm design of the decision making module.

\section{Motion Prediction of CAVs for Decision Making}
In this section, a simplified vehicle kinematic model is proposed for the motion prediction of CAVs. Then, based on this vehicle kinematic model, a discrete motion prediction model is further built for design of the decision-making algorithm.
\subsection{Vehicle Kinematic Model}
Considering that CAVs move with relatively low speeds at the roundabout, in order to reduce the model complexity for algorithm design, the following simplified vehicle kinematic model is applied to the motion prediction of CAVs [35].
\begin{align}
\dot{x}(t)=f[x(t),u(t)]
\end{align}
\begin{align}
f[x(t),u(t)]=
&
\left[
\begin{array}{ccc}
a_x\\
v_x\tan\beta/l_r\\
v_x\cos(\varphi+\beta)/\cos\beta\\
v_x\sin(\varphi+\beta)/\cos\beta\\
\end{array}
\right]
\end{align}
where the state vector $x=[v_x, \varphi, X, Y]^{T}$, and the control vector $u=[a_x, \delta_f]^{T}$. $v_x$ and $\varphi$ are the longitudinal velocity and yaw angle, respectively. $(X, Y)$  is the coordinate position of the vehicle. $a_x$ and $\delta_f$ are the longitudinal acceleration and the steering angle of the front wheel, respectively. The sideslip angle $\beta=\arctan(l_r/(l_f+l_r)\tan\delta_f)$. $l_f$ and $l_r$ are the front and rear wheel bases, respectively.

\subsection{Model-based Motion Prediction}
For algorithm design of motion prediction, the vehicle kinematic model, Eq. (1) is transformed into a time-varying linear system.
\begin{align}
\dot{x}(t)=A_tx(t)+B_tu(t)
\end{align}
where the time-varying coefficient matrices are expressed as
\begin{align}
A_t=\left.\frac{\mathrm{\partial}f}{\mathrm{\partial}x}\right|_{x_t,u_t},\ B_t=\left.\frac{\mathrm{\partial}f}{\mathrm{\partial}u}\right|_{x_t,u_t}
\end{align}

Furthermore, Eq. (3) is discretized as
\begin{align}
\left\{
\begin{array}{lr}
x(k+1)=A_kx(k)+B_ku(k)\\
u(k)=u(k-1)+\Delta{u(k)}\\
\end{array}
\right.
\end{align}
where $x(k)=[v_x(k), \varphi(k), X(k), Y(k)]^{T}$, $A_k=e^{A_t\Delta{T}}$,\ $B_k=\int_{0}^{\Delta{T}}{e^{A_t\tau}}B_td\tau$, $\Delta{T}$ is the sampling time,  $u(k)=[a_x(k), \delta_f(k)]^{T}$, $\Delta{u(k)}=[\Delta{a_x(k)}, \Delta{\delta_f(k)}]^{T}$.

After that, a new state vector is formulated to integrate the original state vector and the control input.
\begin{align}
\xi(k)=[x(k),u(k-1)]^T
\end{align}

As a result, a new discrete state-space form of Eq. (5) is derived as
\begin{align}
\left\{
\begin{array}{lr}
\xi(k+1)=\hat{A}_k\xi(k)+\hat{B}_k\Delta{u(k)}\\
y(k)=\hat{C}_k\xi(k)\\
\end{array}
\right.
\end{align}
where $\hat{A}_k=
\left[
\begin{array}{ccc}
A_k & B_k\\
0_{2\times4} & I_2\\
\end{array}
\right]$,
$\hat{B}_k=
\left[
\begin{array}{ccc}
B_k\\
I_2\\
\end{array}
\right]$, and
$\hat{C}_k=
\left[
\begin{array}{ccc}
I_4 & 0_{4\times2}\\
\end{array}
\right]$.

Then, the predictive horizon $N_p$  and the control horizon $N_c$ are defined, $N_p>N_c$. At the time step $k$, if the state vector $\xi(k)$, the control vector $\Delta{u(k)}$ and coefficient matrices i.e., $\hat{A}_{p,k}$, $\hat{B}_{p,k}$ and $\hat{C}_{p,k}$ are known, the predicted state vectors can be expressed as
\begin{align}
\left\{
\begin{array}{lr}
\xi(p+1|k)=\hat{A}_{p,k}\xi(p|k)+\hat{B}_{p,k}\Delta{u(p|k)}\\
y(p|k)=\hat{C}_{p,k}\xi(p|k)\\
\end{array}
\right.
\end{align}
where $p=k,k+1,\cdot\cdot\cdot,k+N_p-1$.

Supposing that $\hat{A}_{p,k}=\hat{A}_k$, $\hat{B}_{p,k}=\hat{B}_k$ and $\hat{C}_{p,k}=\hat{C}_k$, the state prediction is derived as follows.
\begin{align}
\begin{array}{lr}
\xi(k+1|k)=\hat{A}_k\xi(k|k)+\hat{B}_k\Delta{u(k|k)}\\
\xi(k+2|k)=\hat{A}_k^2\xi(k|k)+\hat{A}_k\hat{B}_k\Delta{u(k|k)}+\hat{B}_k\Delta{u(k+1|k)}\\
\quad \quad\vdots\\
\xi(k+N_c|k)=\hat{A}_k^{N_c}\xi(k|k)+\hat{A}_k^{N_c-1}\hat{B}_k\Delta{u(k|k)}+\cdot\cdot\cdot\\
\quad \quad\quad\quad\quad\quad\quad+\hat{B}_k\Delta{u(k+N_c-1|k)}\\
\quad \quad\vdots\\
\xi(k+N_p|k)=\hat{A}_k^{N_p}\xi(k|k)+\hat{A}_k^{N_p-1}\hat{B}_k\Delta{u(k|k)}+\cdot\cdot\cdot\\
\quad \quad\quad\quad\quad\quad\quad+\hat{A}_k^{N_p-N_c}\hat{B}_k\Delta{u(k+N_c-1|k)}\\
\end{array}
\end{align}

The output vector sequence is defined as
\begin{align}
\Upsilon(k)=[y^T(k+1|k),y^T(k+2|k),\cdot\cdot\cdot,y^T(k+N_p|k)]^T
\end{align}

Based on Eqs. 9 and 10, the predicted motion output vector sequence $\Upsilon(k)$ is expressed as
\begin{align}
\Upsilon(k)=\bar{C}\xi(k|k)+\bar{D}\Delta{\mathbf{u}(k)}
\end{align}
where
$\Delta\mathbf{u}(k)=[\Delta{u}^T(k|k),\Delta{u}^T(k+1|k),\cdot\cdot\cdot,\Delta{u}^T(k+N_c-1|k)]^T$,
$\bar{C}=[(\hat{C}_k\hat{A}_k)^T,(\hat{C}_k\hat{A}_k^2)^T,\cdot\cdot\cdot,(\hat{C}_k\hat{A}_k^{N_p})^T]^T$,
$\bar{D}=
\left[
\begin{array}{ccccc}
\hat{C}_k\hat{B}_k & 0 & 0 & 0\\
\vdots & \vdots & \vdots & \vdots\\
\hat{C}_k\hat{A}_k^{N_c-1}\hat{B}_k & \cdots & \hat{C}_k\hat{A}_k\hat{B}_k & \hat{C}_k\hat{B}_k\\
\vdots & \vdots & \vdots & \vdots\\
\hat{C}_k\hat{A}_k^{N_p-1}\hat{B}_k & \cdots & \hat{C}_k\hat{A}_k^{N_p-N_c+1}\hat{B}_k & \hat{C}_k\hat{A}_k^{N_p-N_c}\hat{B}_k\\
\end{array}
\right]$.
\\

The motion prediction procedure of CAVs is finished. Based on the motion prediction, the control vector sequence $\Delta\mathbf{u}$ for CAVs can be worked out by solving the decision-making problem formulated in the following section.

\section{Algorithm Design of Decision Making using the Game Theoretic Approach}
In this section, two game theoretic approaches, i.e., the Stackelberg game and grand coalition game, are applied to the decision-making problem formulated for the CAVs at the unsignalized roundabout. Firstly, the decision-making payoff function is defined with consideration of safety, comfort and efficiency. Then, multiple constraints related to decision making are proposed. Finally, the game theoretic decision-making issue is optimized with the MPC approach.
\subsection{Payoff Function of Decision Making Considering One Opponent}
As mentioned in Section II, the decision making process of CAVs at the roundabout are divided into three stages, i.e., entering, passing and exiting. The decision making of CAVs at the entering stage can be described as the merging issue from the main road to the round road, and the decision making of CAVs at the passing stage and the exiting stage can be described as the lane-change issue. Correspondingly, two kinds of decision making behaviors of CAVs are defined, i.e., the merging behavior $\alpha$, $\alpha\in\{-1,0,1\}:=$ \{\emph{merging into the inside lane of round road, lane keeping, merging into the outside lane of round road}\}, the lane-change behavior $\beta$, $\beta\in\{-1,0,1\}:=$ \{\emph{left lane change, lane keeping, right lane change}\}. The decision making process of CAVs at the roundabout involves multi-vehicles and multi-lanes, which brings difficulty to the establishment of the decision-making payoff function. In view of this, taking normal lane-change issue as example, the payoff function that considers one opponent (NV) is studied firstly. Then, the decision-making payoff function that deals with complex scenarios at the roundabout is discussed in the next subsection.

In the decision-making payoff function of CAVs, three vital performance indexes are considered including safety, comfort and efficiency. For the HV (CAVi), the payoff function of decision making is defined as
\begin{align}
P^{i}=k_s^{i}P_s^{i}+k_c^{i}P_c^{i}+k_e^{i}P_e^{i}
\end{align}
where $P_s^{i}$, $P_c^{i}$ and $P_e^{i}$ denote the payoffs of safety, comfort and efficiency, respectively. $k_s^{i}$, $k_c^{i}$ and $k_e^{i}$ are the weighting coefficients, which reflect the driving style of CAVi. Based on the analysis of human drivers' driving behaviors [31,35], the weighting coefficients of the three different driving styles are listed in Table I.

The payoff of driving safety  $P_s^{i}$ includes the longitudinal safety, lateral safety and lane keeping safety, which is expressed as
\begin{align}
P_s^{i}=(1-(\beta^{i})^2)P_{s-log}^{i}+(\beta^{i})^2P_{s-lat}^{i}+(1-(\beta^{i})^2)P_{s-lk}^{i}
\end{align}
where $P_{s-log}^{i}$, $P_{s-lat}^{i}$ and $P_{s-lk}^{i}$ denote the payoffs of the longitudinal, lateral and lane keeping safety, respectively.

\begin{table}[t]
	\renewcommand{\arraystretch}{1.0}
	\caption{Weighting Coefficients of Different Driving Styles}
\setlength{\tabcolsep}{6mm}
	\centering
	\label{table_1}
	\resizebox{\columnwidth}{!}{
		\begin{tabular}{l l l l}
			\hline\hline \\[-3mm]
			\multirow{2}{*}{Driving Characteristic} &\multicolumn{3}{c}{Weighting Coefficients} \\
\cline{2-4} & $k_s^{i}$ & $k_c^{i}$ & $k_e^{i}$ \\
\hline
			\multicolumn{1}{c}{Aggressive}  & 0.4 & $ 0.5 $ & 0.1\\
			\multicolumn{1}{c}{Normal} & 0.3 & 0.3 & 0.4 \\
\multicolumn{1}{c}{Conservative} & 0.1 & 0.4 & 0.5 \\
			\hline\hline
		\end{tabular}
	}
\end{table}

The payoff of the longitudinal safety $P_{s-log}^{i}$, which is related to the longitudinal gap and relative velocity with respective to LV, is defined by
\begin{align}
\begin{array}{lr}
P_{s-log}^{i}=k_{v-log}^{i}[(\Delta v_{x,log}^{i})^2+\varepsilon]^{\eta_{log}^{i}}
+k_{s-log}^{i}(\Delta s_{log}^{i})^2
\tag{14a}
\end{array}
\end{align}
\begin{align}
\Delta v_{x,log}^{i}=v_{x}^{LV}-v_{x}^{i}
\tag{14b}
\end{align}
\begin{align}
\begin{array}{lr}
\Delta s_{log}^{i}=[(X^{LV}-X^{i})^2+(Y^{LV}-Y^{AVi})^2]^{1/2}-L_{v}
\tag{14c}
\end{array}
\end{align}
\begin{align}
\eta_{log}^{i}=\mathrm{sgn}(\Delta v_{x,log}^{i})
\tag{14d}
\end{align}
where $v_{x}^{LV}$ and $v_{x}^{i}$ denote the longitudinal velocities of LV and CAVi, respectively. $(X^{LV},Y^{LV})$  and $(X^{i},Y^{i})$ are the positions of LV and CAVi, respectively. $k_{v-log}^{i}$ and $k_{s-log}^{i}$ are the weighting coefficients. $\Delta v_{x,log}^{i}$ and $\Delta s_{log}^{i}$ are the relative velocity and longitudinal gap between LV and CAVi. $L_{v}$ is a safety coefficient that takes into consideration the length of the vehicle. $\varepsilon$ is a design parameter of a small positive value to avoid zero denominator in the calculation.

The payoff of the lateral safety $P_{s-lat}^{i}$, which is related to the relative distance and relative velocity with respective to NV, is expressed as
\begin{align}
\begin{array}{lr}
P_{s-lat}^{i}=k_{v-lat}^{i}[(\Delta v_{x,lat}^{i})^2+\varepsilon]^{\eta_{lat}^{i}}+k_{s-lat}^{i}(\Delta s_{lat}^{i})^2
\tag{15a}
\end{array}
\end{align}
\begin{align}
\Delta v_{x,lat}^{i}=v_{x}^{i}-v_{x}^{NV}
\tag{15b}
\end{align}
\begin{align}
\begin{array}{lr}
\Delta s_{lat}^{i}=[(X^{i}-X^{NV})^2+(Y^{i}-Y^{NV})^2]^{1/2}-L_{v}
\tag{15c}
\end{array}
\end{align}
\begin{align}
\eta_{lat}^{i}=\mathrm{sgn}(\Delta v_{x,lat}^{i})
\tag{15d}
\end{align}
where $v_{x}^{NV}$ is the longitudinal velocity of NV. $(X^{NV},Y^{NV})$ is the position of NV. $k_{v-lat}^{i}$ and $k_{s-lat}^{i}$  are the weighting coefficients. $\Delta v_{x,lat}^{i}$ and $\Delta s_{lat}^{i}$ are the relative velocity and longitudinal gap between CAVi and NV.

The payoff of the lane keeping safety $P_{s-lk}^{i}$ is associated with the lateral distance error and the yaw angle error between the predicted position of CAVi and the center line of the lane, which is written as
\begin{align}
P_{s-lk}^{i}=k_{y-lk}^{i}/[(\Delta y^{i})^2+\varepsilon]+k_{\varphi-lk}^{i}/[(\Delta \varphi^{i})^2+\varepsilon]
\tag{16}
\end{align}
where $\Delta y^{i}$ and $\Delta \varphi^{i}$  are the lateral distance error and yaw angle error, $k_{y-lk}^{i}$  and $k_{\varphi-lk}^{i}$  are the weighting coefficients.

The payoff of the ride comfort  $P_{c}^{i}$, which is associated with the acceleration of CAVi, is expressed as
\begin{align}
\begin{array}{lr}
P_{c}^{i}=k_{a_x}^{i}/[(a_{x}^{i})^2+\varepsilon]+k_{a_y}^{i}/[(a_{y}^{i})^2+\varepsilon]
\tag{17}
\end{array}
\end{align}
where $a_{x}^{i}$  and  $a_{y}^{i}$ are the longitudinal and lateral accelerations of CAVi, $k_{a_x}^{i}$ and $k_{a_y}^{i}$ are the weighting coefficients.

In addition, the payoff of travel efficiency $P_e^{i}$  is defined as a function related to the longitudinal velocity of CAVi, which is written as
\begin{align}
P_e^{i}=k_e^{i}/[(v_{x}^{i}-v_{x}^{\mathrm{max}})^2+\varepsilon]
\tag{18}
\end{align}
where $v_{x}^{\mathrm{max}}$  is the maximum velocity, and $k_e^{i}$  is the weighting coefficient.

\subsection{Payoff Function of Decision Making Considering Multiple Opponents}
After analysis of the aforementioned expression of payoff function, it can be found that the payoff of safety is directly influenced by the decision-making behavior, and the effect on the payoffs of comfort and efficiency is indirect. Additionally, the payoffs of comfort and efficiency are only related to the motion state of CAVi itself. However, the payoff of safety is also related to the motion states and driving behaviors of surrounding vehicles. Therefore, the formulation of the payoff function, which considers multiple opponents (more than one NV), focuses on the definition of the payoff of safety.

Considering the multi-vehicle and multi-lane scenario at the roundabout, the payoff function of safety for CAVi is defined by
\begin{align}
P_s^{i}=\Gamma(P_{s-log}^{i}, P_{s-lat}^{i}, P_{s-lk}^{i}, \alpha^{i}, \beta^{i})
\tag{19}
\end{align}
where the function $\Gamma$ is different at the three decision-making stages.

At the entering stage, for CAVi, only the merging behavior need to be considered. Therefore, $\Gamma$ is derived as
\begin{align}
\Gamma=(1-(\alpha^{i})^2)P_{s-log}^{i}+(\alpha^{i})^2P_{s-lat}^{i}+(1-(\alpha^{i})^2)P_{s-lk}^{i}
\tag{20}
\end{align}

The payoff of lateral safety is associated with three NVs, which is expressed as
\begin{align}
\begin{array}{lr}
P_{s-lat}^{i}=0.25(\alpha^{i}+1)^2P_{s-lat}^{i\&NV1}+P_{s-lat}^{i\&NV2}\\[1.4ex]
\quad \quad \quad \quad \quad +0.25(\alpha^{i}-1)^2P_{s-lat}^{i\&NV3}
\end{array}
\tag{21}
\end{align}
where $P_{s-lat}^{i\&NV1}$, $P_{s-lat}^{i\&NV2}$ and $P_{s-lat}^{i\&NV3}$ are the payoffs of lateral safety associated with NV1, NV2 and NV3, respectively.

At the passing and exiting stage, $\Gamma$ is only related to the lane-change behavior, which is written as
\begin{align}
\Gamma=(1-(\beta^{i})^2)P_{s-log}^{i}+P_{s-lat}^{i}+(1-(\beta^{i})^2)P_{s-lk}^{i}
\tag{22}
\end{align}

From Fig. 2, it can be found that the payoff of lateral safety is not only associated with the NVs, but also influenced by the merging behavior of NV2. As a result, $P_{s-lat}^{i}$ is expressed as
\begin{align}
P_{s-lat}^{i}=(\beta^{i})^2P_{s-lat}^{i\&NV1}+(\alpha^{NV2})^2P_{s-lat}^{i\&NV2}
\tag{23}
\end{align}
where $\alpha^{NV2}$ is the merging behavior of NV2.

\subsection{Constraints of Decision Making}
As to the safety, comfort and efficiency of CAVi in the process of decision making, some constraints must be considered. The safety constraints for CAVi are defined as
\begin{align}
|\Delta s^{i}|\leq\Delta s^{\mathrm{max}},|\Delta y^{i}|\leq\Delta y^{\mathrm{max}},|\Delta \varphi^{i}|\leq\Delta \varphi^{\mathrm{max}}
\tag{24}
\end{align}

The constraints for ride comfort are given by
\begin{align}
|a_{x}^{i}|\leq a_x^{\mathrm{max}},|a_{y}^{i}|\leq a_y^{\mathrm{max}}
\tag{25}
\end{align}

Moreover, the constraint for travel efficiency is represented as
\begin{align}
|v_{x}^{i}|\leq v_{x}^{\mathrm{max}}
\tag{26}
\end{align}

Additionally, the control constraint of $\Delta a_x^{i}$ is defined as
\begin{align}
|\Delta a_x^{i}|\leq\Delta a_x^{\mathrm{max}}
\tag{27}
\end{align}

The control constraints of $\Delta \delta_f^{i}$ and $\delta_f^{i}$ are given by
\begin{align}
|\Delta \delta_f^{i}|\leq\Delta \delta_f^{\mathrm{max}},  |\delta_f^{i}|\leq \delta_f^{\mathrm{max}}
\tag{28}
\end{align}

Finally, the aforementioned constraints for CAVi can be expressed in a compact form as
\begin{align}
\begin{array}{lr}
\Xi^{i}(\Delta s^{i},\Delta y^{i},\Delta \varphi^{i},a_{x}^{i},a_{y}^{i}, v_{x}^{i},\Delta a_x^{i},\Delta \delta_f^{i},\delta_f^{i})\leq0
\tag{29}
\end{array}
\end{align}

The value settings of the constraint boundaries are listed in Table II.
\begin{table}[h]
	\renewcommand{\arraystretch}{1.3}
	\caption{Constraint Boundaries for Decision Making}
\setlength{\tabcolsep}{6mm}
	\centering
	\label{table_2}
	\resizebox{\columnwidth}{!}{
		\begin{tabular}{c c | c c}
			\hline\hline \\[-4mm]
            Parameter & Value & Parameter & Value \\
\hline
			$\Delta s^{\mathrm{max}}$/ $\mathrm{(m)}$  & 0.8 & $v_{x}^{\mathrm{max}}$/ $\mathrm{(m/s)}$ & 30\\
           $\Delta y^{\mathrm{max}}$/ $\mathrm{(m)}$  & 0.2 & $\Delta a_x^{\mathrm{max}} $/ $\mathrm{(m/s^2)}$ & 0.1\\
           $\Delta \varphi^{\mathrm{max}}$/ $\mathrm{(deg)}$  & 2 & $\Delta \delta_f^{\mathrm{max}} $/ $\mathrm{(deg)}$ & 0.3\\
           $a_x^{\mathrm{max}}$/ $\mathrm{(m/s^2)}$ & 8 & $\delta_f^{\mathrm{max}} $/ $\mathrm{(deg)}$ & 30\\
           $a_y^{\mathrm{max}}$/ $\mathrm{(m/s^2)}$& 5 & -- & --\\
			\hline\hline
		\end{tabular}
	}
\end{table}

\subsection{Decision Making with the Game Theory and MPC Optimization}
As Fig. 4 shows, two game theoretic approaches, i.e., the Stackelberg game and grand coalition game, are utilized to address the decision-making issue of CAVs at the unsignalized roundabout. In the Stackelberg game, there exits a leader and some followers. All the players try to maximize the individual payoff. Differing from the normal noncooperative game, the leader is endowed the power of predicting follower's strategy given its own. Namely, the decision-making results of followers have significant effects on the leader's decision making [37]. In the grand coalition game, all the players form a large group that aims to maximize the payoff of the grand coalition [38]. The grand coalition game is a typical cooperative game.

\begin{figure}[h]\centering
	\includegraphics[width=7cm]{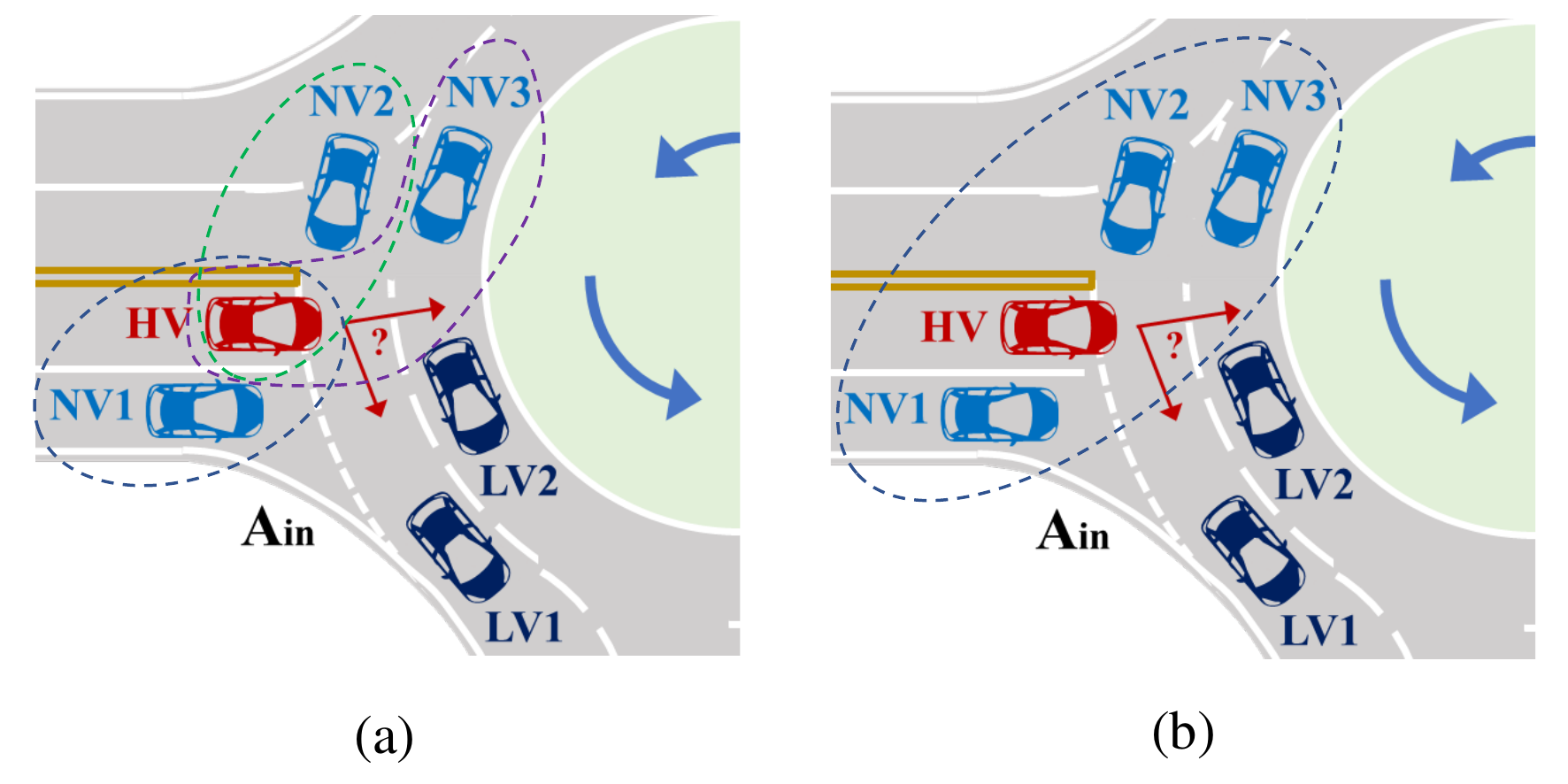}
	\caption{Decision making of CAVs at the roundabout with the game theoretic approach: (a) The Stackelberg game approach; (b) The grand coalition game approach.}\label{FIG_4}
\end{figure}

Above all, the decision-making vector of CAVi is defined as $\hat{u}^{i}=[\Delta a_x^{i},\Delta \delta_f^{i},\alpha^{i}, \beta ^{i}]^T$. Based on the established payoff function of CAVi, the game theoretic-based decision making algorithm is designed as follows.

(1) Stackelberg game approach
\begin{align}
\begin{array}{lr}
\hat{u}^{HV*}=\arg\max\limits_{\hat{u}^{HV}\in U^{HV}}(\min \limits_{\Lambda}P^{HV}(\hat{u}^{HV},\hat{u}^{NV1},\\[1.4ex]
\quad \quad \quad \quad \quad \quad \quad \quad \hat{u}^{NV2},\hat{u}^{NV3}))
\end{array}
\tag{30a}
\end{align}
\begin{align}
\begin{array}{lr}
\Lambda\triangleq\{\hat{u}^{NV1}\in U^{NV1*},\hat{u}^{NV2}\in U^{NV2*},\\[1.4ex]
\quad \quad \quad \quad \quad \quad   \hat{u}^{NV3}\in U^{NV3*}\}
\end{array}
\tag{30b}
\end{align}
\begin{align}
\begin{array}{lr}
U^{NV1*}(\hat{u}^{HV})\triangleq\{\hat{u}^{NV1*}\in U^{NV1}: P^{NV1}(\hat{u}^{HV},\\[1.4ex]
\ \hat{u}^{NV1*}) \geq P^{NV1}(\hat{u}^{HV},\hat{u}^{NV1}),\forall \hat{u}^{NV1}\in U^{NV1} \}
\end{array}
\tag{30c}
\end{align}
\begin{align}
\begin{array}{lr}
U^{NV2*}(\hat{u}^{HV})\triangleq\{\hat{u}^{NV2*}\in U^{NV2}: P^{NV1}(\hat{u}^{HV},\\[1.4ex]
\ \hat{u}^{NV2*}) \geq P^{NV2}(\hat{u}^{HV},\hat{u}^{NV2}),\forall \hat{u}^{NV2}\in U^{NV2} \}
\end{array}
\tag{30d}
\end{align}
\begin{align}
\begin{array}{lr}
U^{NV3*}(\hat{u}^{HV})\triangleq\{\hat{u}^{NV3*}\in U^{NV3}: P^{NV3}(\hat{u}^{HV},\\[1.4ex]
\ \hat{u}^{NV3*})\geq P^{NV3}(\hat{u}^{HV},\hat{u}^{NV3}),\forall \hat{u}^{NV3}\in U^{NV3} \}
\end{array}
\tag{30e}
\end{align}
s.t. $\Xi^{HV}\leq0$, $\Xi^{NV1}\leq0$, $\Xi^{NV2}\leq0$, $\Xi^{NV3}\leq0$. \\
where $\hat{u}^{HV*},\hat{u}^{NV1*},\hat{u}^{NV2*}$ and $\,\hat{u}^{NV3*}$  denote the optimal decision-making results of HV, NV1, NV2 and NV3, respectively.

(2) Grand coalition game approach
\begin{align}
\begin{array}{lr}
(\hat{u}^{HV*},\hat{u}^{NV1*},\hat{u}^{NV2*},\hat{u}^{NV3*})=\arg\max(\omega^{HV} P^{HV}\\[1.4ex]
\quad \quad \quad \quad +\omega^{NV1}P^{NV1}+\omega^{NV2}P^{NV2}+\omega^{NV3}P^{NV3})
\end{array}
\tag{31}
\end{align}
s.t. $\Xi^{HV}\leq0$, $\Xi^{NV1}\leq0$, $\Xi^{NV2}\leq0$, $\Xi^{NV3}\leq0$.\\
where $\omega^{HV}$, $\omega^{NV1}$, $\omega^{NV2}$ and $\omega^{NV3}$ are the allocation coefficients of HV, NV1, NV2 and NV3, respectively. In this paper, the total payoff is equally allocated to the CAVs.

Based on the prediction information of motion states, the decision-making algorithm for CAVs can supply the low-level motion planning and control modules with accurate and reliable decisions. In terms of this, the MPC approach is applied to the optimization problem formulated for decision-making.

According to the motion prediction algorithm proposed in Section III, the payoff function sequence for CAVi at the time step k is expressed as
\begin{align}
P^{i}(k+1|k),P^{i}(k+2|k),\cdots,P^{i}(k+N_p|k)
\tag{32}
\end{align}

To transform the decision-making issue of CAVs at the roundabout into a MPC optimization, the following cost function is established.
\begin{align}
J^{i}=1/(P^{i}+\varepsilon)
\tag{33}
\end{align}

\begin{figure*}[!t]\centering
	\includegraphics[width=18cm]{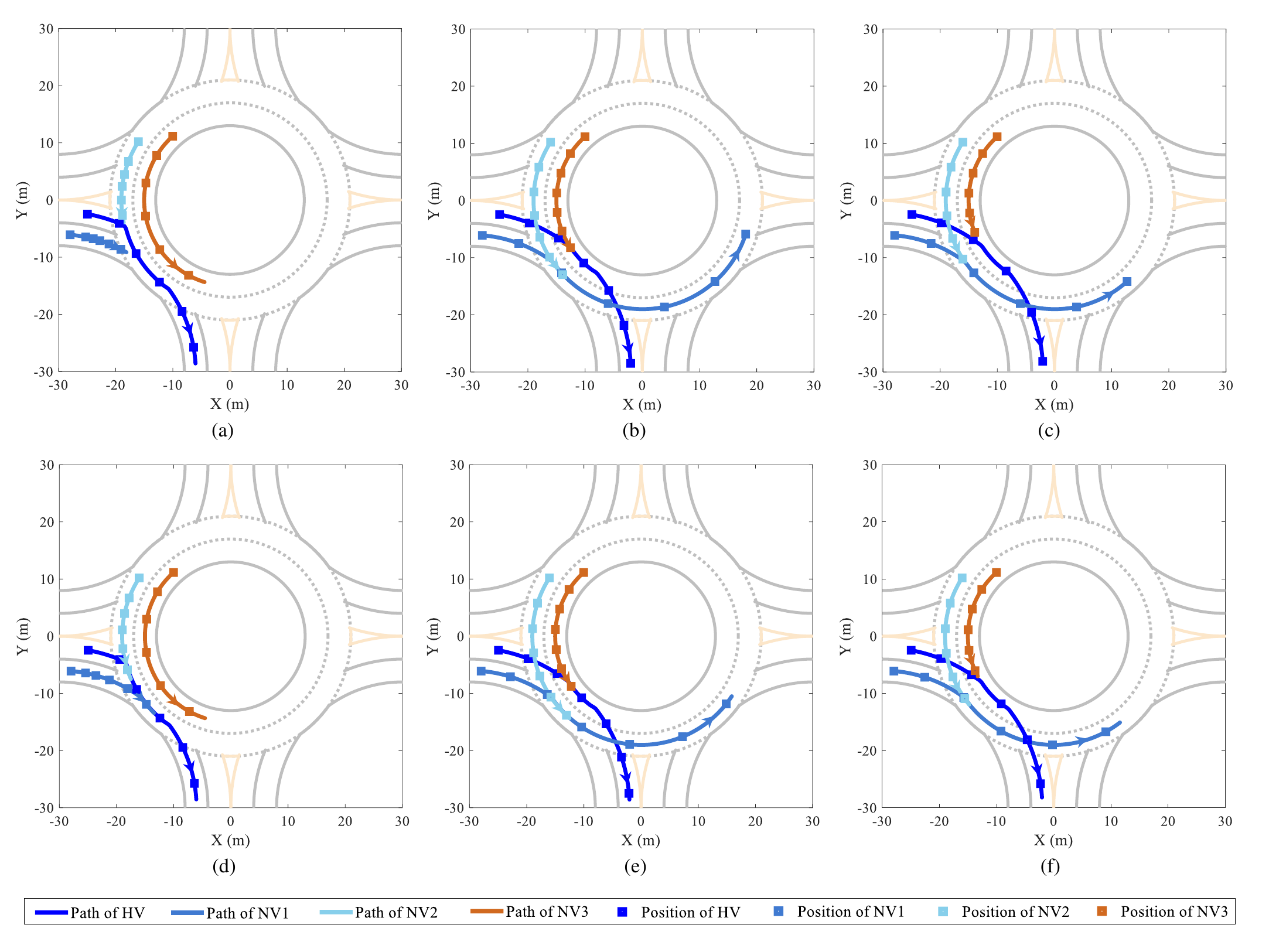}
	\caption{Decision-making results of CAVs in Case 1: (a) SG approach in Scenario A; (b) SG approach in Scenario B; (c) SG approach in Scenario C; (d) GC approach in Scenario A; (e) GC approach in Scenario B; (f) GC approach in Scenario C.}\label{FIG_5}
\end{figure*}

The decision-making sequence of CAVi is written as
\begin{align}
\hat{u}^{i}(k|k),\hat{u}^{i}(k+1|k),\cdots,\hat{u}^{i}(k+N_c-1|k)
\tag{34}
\end{align}
where $\hat{u}^{i}(q|k)=[\Delta a_x^{i}(q|k),\Delta \delta_f^{i}(q|k),\alpha^{i}(q|k), \beta ^{i}(q|k)]^T$, $q=k,k+1,\cdots,k+N_c-1$.

Furthermore, the performance function of CAVi for decision making is defined as
\begin{align}
\Pi^{i}=\sum\limits_{p=k+1}^{k+N_p}||J^{i}(p|k)||_Q^2+\sum\limits_{q=k}^{k+N_c-1}||\hat{u}^{i}(q|k)||_R^2
\tag{35}
\end{align}
where $Q$ and $R$ are the weighting matrices.

Combined with the MPC optimization, the Stackelberg game approach for decision making is transformed into
\begin{align}
\begin{array}{lr}
(\hat{\mathbf{u}}^{HV*},\hat{\mathbf{u}}^{NV1*},\hat{\mathbf{u}}^{NV2*},\hat{\mathbf{u}}^{NV3*})=\arg\min\Pi^{HV}\\[1.4ex]
\mathrm{s.t.}\quad  \min\Pi^{NV1}, \min\Pi^{NV2}, \min\Pi^{NV3},\\[1.4ex]
\quad \Xi^{HV}\leq0, \Xi^{NV1}\leq0, \Xi^{NV2}\leq0, \Xi^{NV3}\leq0.
\end{array}
\tag{36}
\end{align}
where $\hat{\mathbf{u}}^{HV*}$, $\hat{\mathbf{u}}^{NV1*}$, $\hat{\mathbf{u}}^{NV2*}$, and $\hat{\mathbf{u}}^{NV3*}$  denote the optimal decision-making sequences of HV, NV1, NV2 and NV3, respectively.

Combined with the MPC optimization, the grand coalition game approach for decision making is expressed as
\begin{align}
\begin{array}{lr}
(\hat{\mathbf{u}}^{HV*},\hat{\mathbf{u}}^{NV1*},\hat{\mathbf{u}}^{NV2*},\hat{\mathbf{u}}^{NV3*})=\arg\min[\omega^{HV}\Pi^{HV}\\[1.4ex]
\quad\quad\quad+\omega^{NV1}\Pi^{NV1}+\omega^{NV2}\Pi^{NV2}+\omega^{NV3}\Pi^{NV3}]\\[1.4ex]
\mathrm{s.t.} \quad \Xi^{HV}\leq0, \Xi^{NV1}\leq0, \Xi^{NV2}\leq0, \Xi^{NV3}\leq0.
\end{array}
\tag{37}
\end{align}

In the decision-making algorithm, three NVs are discussed, which is adequate for the two-lane unsignalized roundabout. If there exits more lanes at the roundabout, the proposed algorithm can be advanced easily.

\section{Testing Results and Analysis}
To verify the feasibility and effectiveness of the designed decision-making algorithm, three testing cases of the unsignalized roundabout are designed and carried out in this section. All the driving scenarios are established and implemented on the MATLAB/Simulink platform.

\subsection{Testing Case 1}
Case 1 focuses on the decision-making issue of entering the roundabout. HV on the inside lane of the main road $M1$ wants to enter the round road $RR$ from the entrance $A_{in}$ and exit from $B_{out}$ to the main road $\hat{M}2$ . At the entering stage, HV must interact with three NVs, i.e., NV1 on the outside lane of the main road $M1$, NV2 on the outside lane of the round road $RR$ and NV3 on the inside lane of the round road $RR$, and then make the optimal decision.

\begin{figure}[!t]\centering
	\includegraphics[width=7cm]{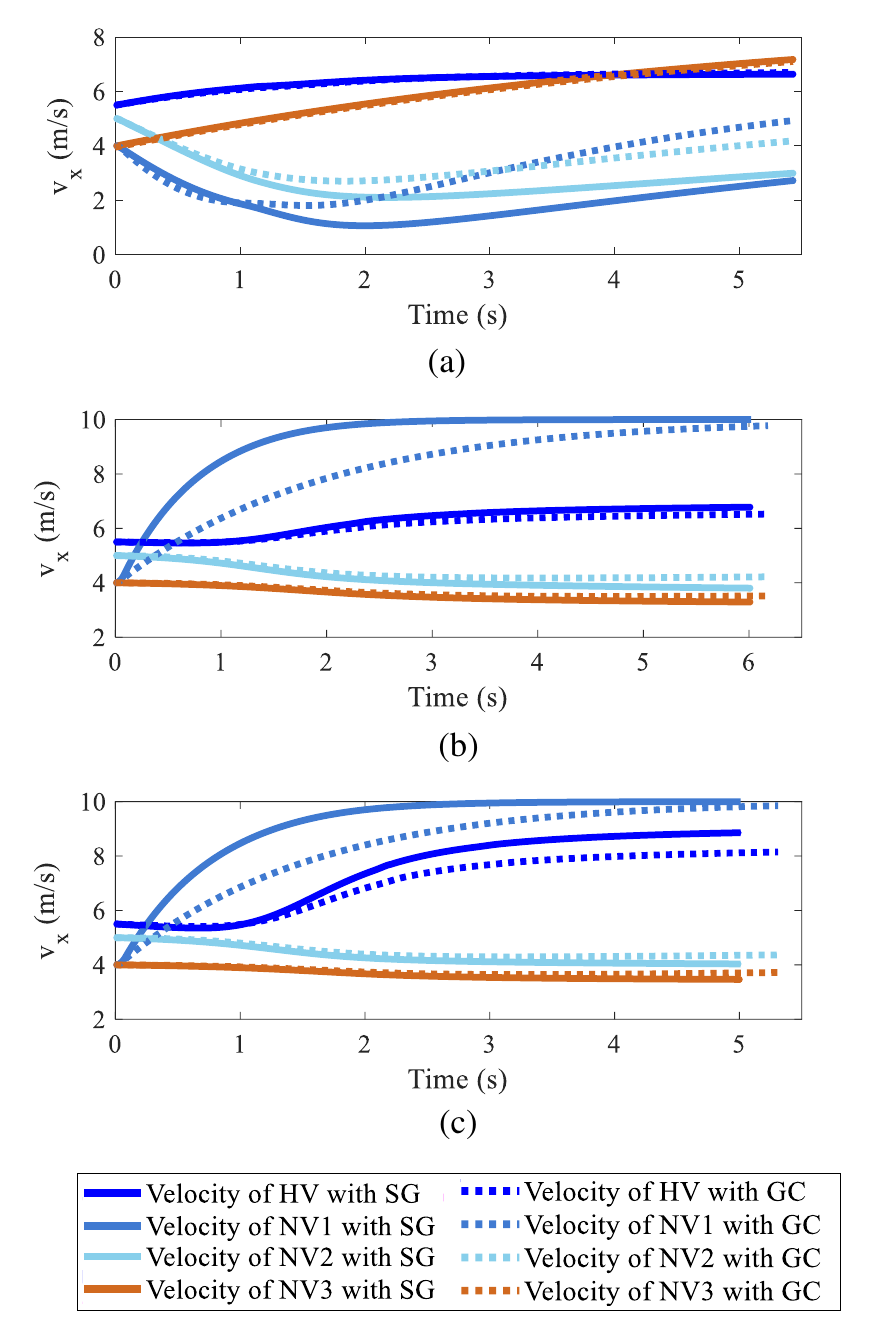}
	\caption{Velocities of CAVs in Case 1: (a) Scenario A; (b) Scenario B; (c) Scenario C.}\label{FIG_6}
\end{figure}

\begin{figure}[!t]\centering
	\includegraphics[width=8.5cm]{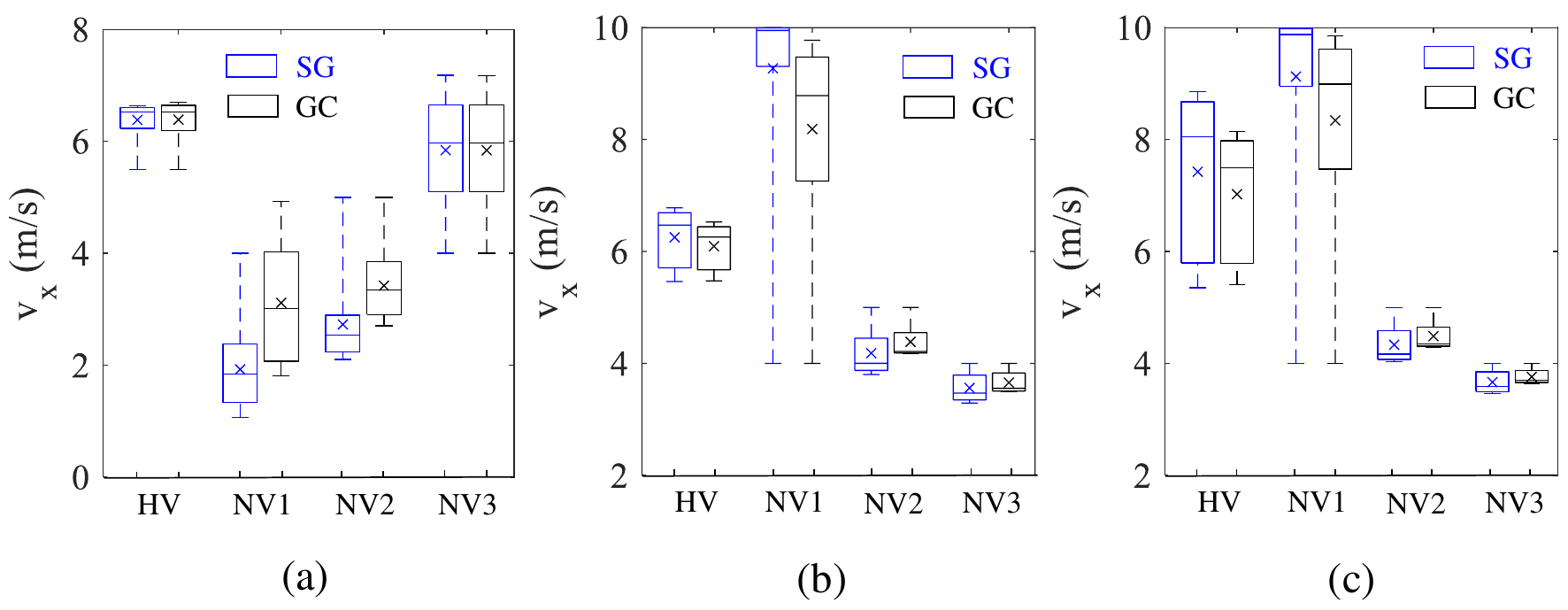}
	\caption{Box plots of velocities in Case 1: (a) Scenario A; (b) Scenario B; (c) Scenario C.}\label{FIG_7}
\end{figure}

\begin{figure}[!t]\centering
	\includegraphics[width=8.5cm]{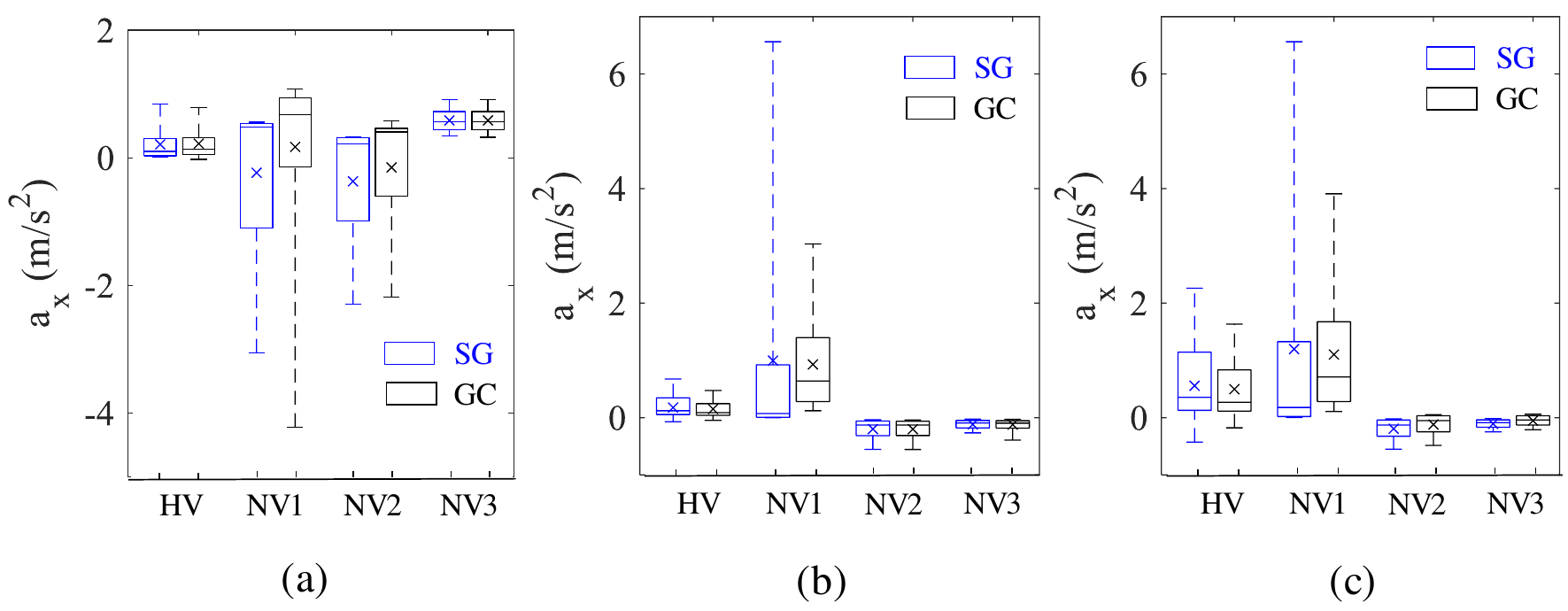}
	\caption{Box plots of longitudinal accelerations in Case 1: (a) Scenario A; (b) Scenario B; (c) Scenario C.}\label{FIG_8}
\end{figure}

\begin{figure}[!t]\centering
	\includegraphics[width=8.5cm]{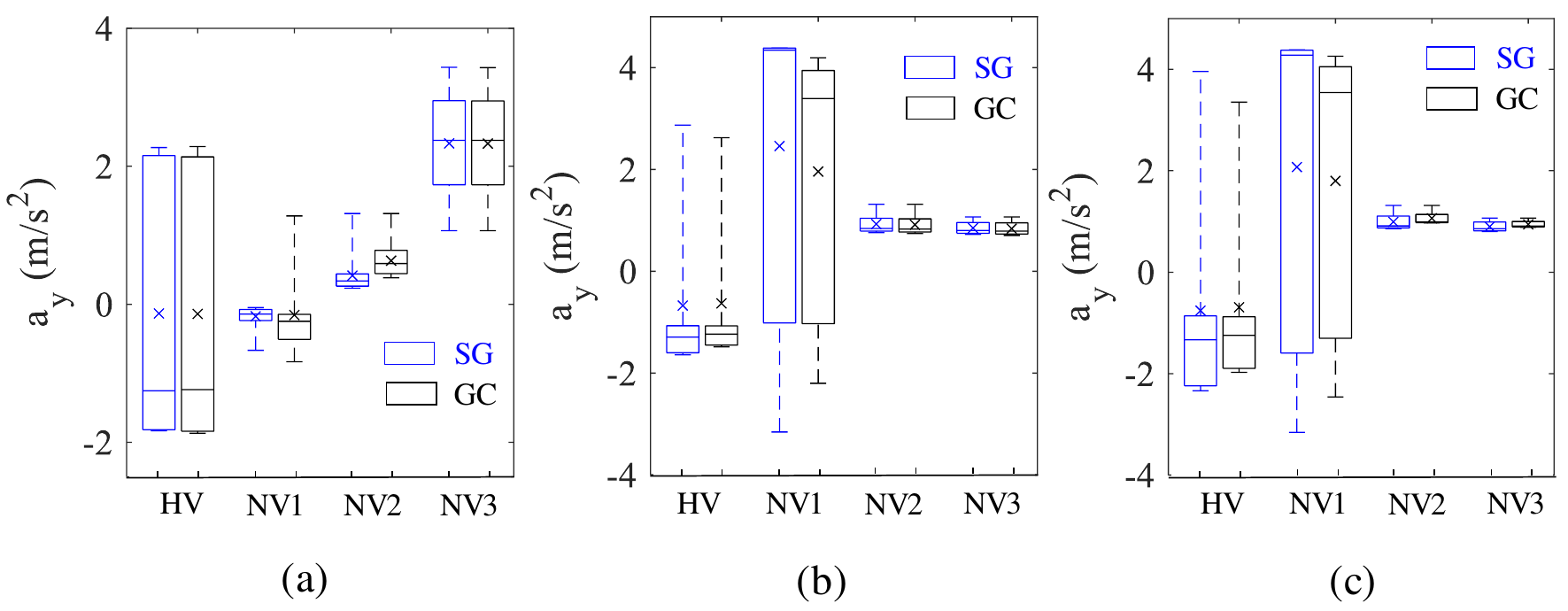}
	\caption{Box plots of lateral accelerations in Case 1: (a) Scenario A; (b) Scenario B; (c) Scenario C.}\label{FIG_9}
\end{figure}

\begin{figure}[!t]\centering
	\includegraphics[width=8.5cm]{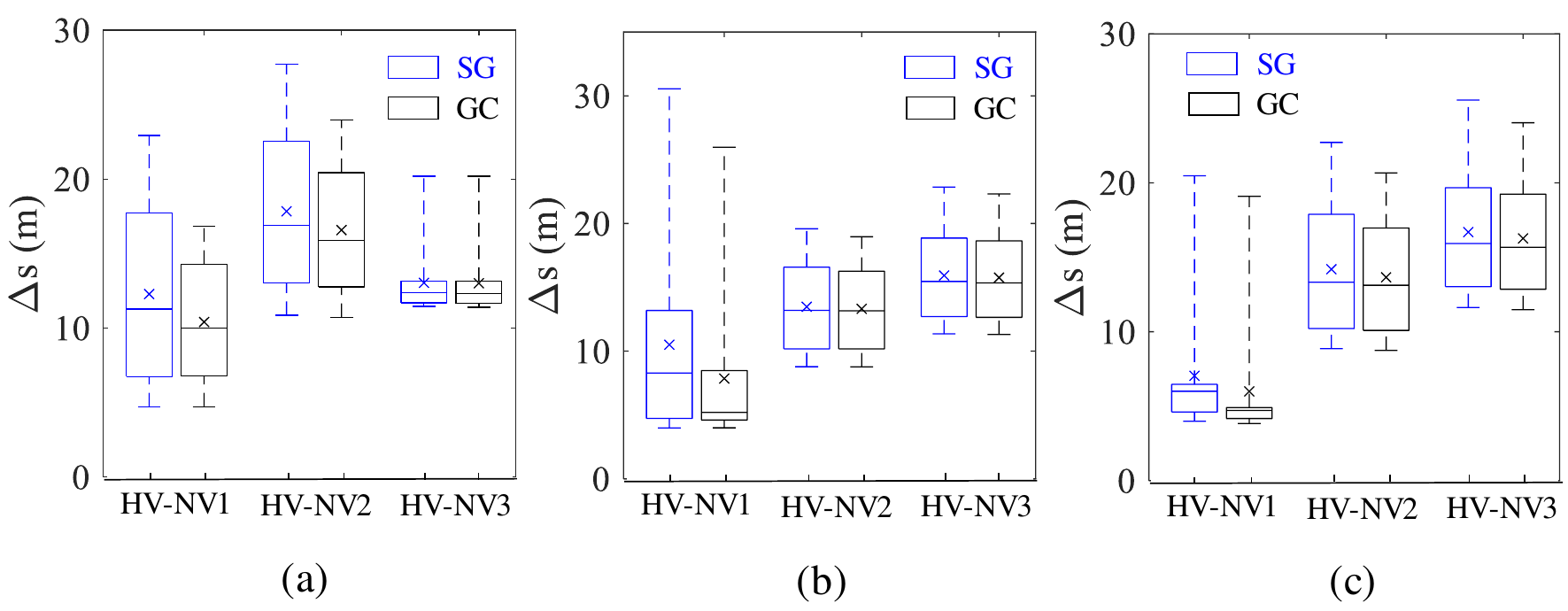}
	\caption{Box plots of relative distances between HV and other NVs in Case 1: (a) Scenario A; (b) Scenario B; (c) Scenario C.}\label{FIG_10}
\end{figure}

To study the effect of personalized driving on decision making, three typical scenarios are designed in this case. In Scenario A, the driving styles of HV, NV1, NV2 and NV3 are normal, conservative, normal and normal, respectively. In Scenario B, the driving styles of the four CAVs are normal, aggressive, normal and normal, respectively. In Scenario C, the driving styles of the four CAVs are aggressive, aggressive, normal and normal, respectively. In this case, the initial position coordinates of HV, NV1, NV2, NV3, LV1 and LV2 are set as (-25, -2.45), (-28, -6.08), (-16, 10.25), (-10, 11.18), (-16, -10.25) and (-14, -5.38), respectively. In addition, the initial velocities of HV, NV1, NV2, NV3, LV1 and LV2 are set to be 5.5 m/s, 4 m/s, 5 m/s, 4 m/s, 8 m/s, 8 m/s, respectively. In this case, the performances of two game theoretic decision making approaches, i.e., Stackelberg game (SG) and grand coalition game (GC), are tested and verified. The testing results are illustrated in Figs. 5-10.

\begin{figure*}[!t]\centering
	\includegraphics[width=18cm]{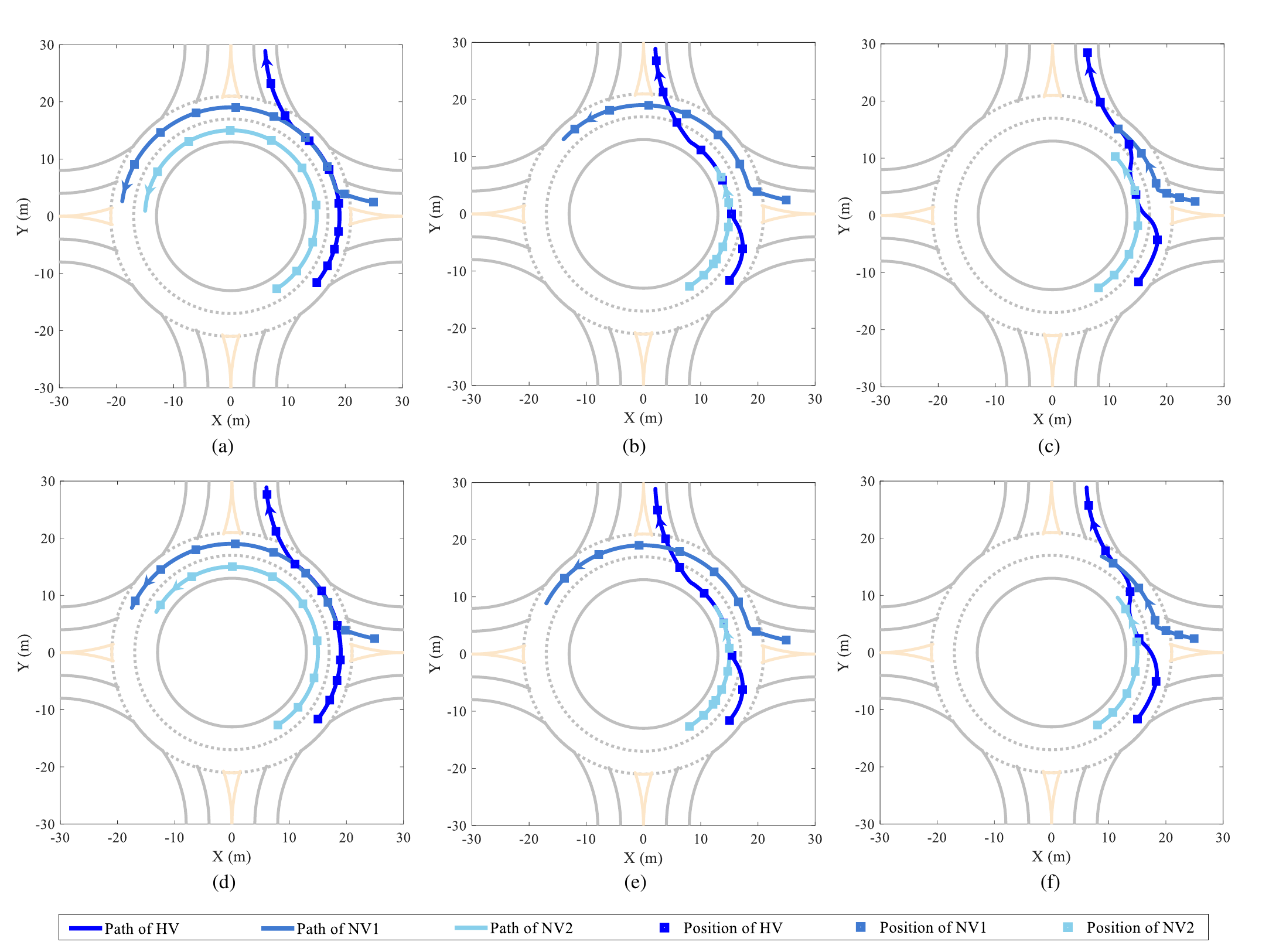}
	\caption{Decision-making results of CAVs in Case 2: (a) SG approach in Scenario A; (b) SG approach in Scenario B; (c) SG approach in Scenario C; (d) GC approach in Scenario A; (e) GC approach in Scenario B; (f) GC approach in Scenario C.}\label{FIG_11}
\end{figure*}

Fig. 5 shows the decision-making results of CAVs in Case 1. It can be found that different driving styles of NVs leads to different decision-making results of HV. In Scenario A, due to the conservative driving style of NV1, NV1 cares more about driving safety and give way for HV. As a result, HV prefers to merge into the outside lane of the round road $RR$. In Scenario B, the driving style of NV1 is aggressive, which causes the aggressiveness of NV1 in fighting for the right of way. Finally, HV decides to merge into the inside lane of the round road $RR$. In the above two scenarios, HV is set as the normal driving style. In Scenario C, both HV and NV1 are the aggressive driving styles. Although it yields to the same decision-making result of HV, the greater concessions of NV2 and NV3 can been seen in Fig. 5. In addition, it can be seen that both SG and GC lead to the similar decision-making results of HV in merging behaviors. However, there exit obvious differences in driving velocity. Figs 6 and 7 shows the change and distribution of vehicle velocities in Case 1. The aggressive driving style brings larger driving velocity due to the priority of travel efficiency. However, to guarantee the driving safety, the conservative driving style results in slower velocity. Moreover, it can also be found that the effect of driving style on decision making is weakened in the GC approach. The vehicle velocity of the aggressive driving style is decreased and the vehicle velocity of the conservative driving style is increased in the GC approach compared with the SG approach. SG belongs to the non-cooperative game approach, in which each player prefers to minimize the individual cost. However, GC is a typical cooperative game approach, which aims to minimize the cost of the whole grand coalition. Therefore, the vehicle velocity of the conservative driving style is increased in GC, which indicates the travel efficiency of the whole traffic system is increased, but the personalized driving is sacrificed.

Figs. 8 and 9 show the distribution of longitudinal and lateral accelerations in this case, from which we can see that vehicles with the aggressive driving styles have largest acceleration in the three driving styles due to pursuing the travel efficiency. To guarantee the driving safety, the conservative driving style leads to great deceleration. The relative distances between HV and other NVs are illustrated in Fig. 10. If two vehicles are aggressive, it will result in the smallest safe distance, which increases the risk of collision. On the contrary, conservative driving style brings larger safe distance. Additionally, whether the acceleration or the relative distance, the similar conclusion can be drawn that the personalized driving characteristic is weakened in the GC approach.

\subsection{Testing Case 2}
Case 2 mainly deals with the decision-making issue that passing and exiting the roundabout. HV on the outside lane of the round road $RR$ wants to exit from $D_{out}$ to the main road $\hat{M}4$ . At the passing stage, HV must address the merging of NV1 from the main road $M3$. Therefore, HV has to interact with NV1 and NV2 on the inside lane of the round road $RR$, and then make the decision, i.e., slowing down and giving way for NV1, speeding up and fighting for the right of way with NV1, or changing lanes to the inside lane of the round road $RR$.

This case studies the effects of personalized driving on decision making as well. Then, three different scenarios are designed in this case. In Scenario A, the driving styles of HV, NV1 and NV2 are conservative, normal and normal, respectively. In Scenario B, the driving styles of the three CAVs are all normal. In Scenario C, the driving styles of the three CAVs are aggressive, normal and normal, respectively. In this case, the initial position coordinates of HV, NV1, NV2, LV1 and LV2 are set as (15, -11.66), (25, 2.45), (8, -12.68), (15, 11.66) and (13, 7.48), respectively. In addition, the initial velocities of HV, NV1, NV2, LV1 and LV2 are set to be 5.5 m/s, 5 m/s, 4 m/s, 8 m/s, 5 m/s, respectively. The performances of the two game theoretic decision making approaches are evaluated and analyzed as well. The testing results are displayed in Figs. 11-16.

The decision-making results of CAVs in Case 2 are displayed in Fig. 11. It can be found that, in Scenario A, HV chooses lane keeping and gives way for NV1 due to the conservative driving style. However, both in Scenario B and Scenario C, the lane-change decision is made by HV. Especially in Scenario C, the driving style of HV is aggressive. To maximize the travel efficiency, a double lane-change behavior is conducted due to the lower driving velocity of NV1 and LV2. It can be concluded that different driving styles of HV lead to different decision-making results, which reflects the personalized driving preferences. Two game theoretic decision making approaches show differences in the decision making of driving velocity, which is illustrated in Figs. 12 and 13. GC tends to increase the velocity of the conservative vehicle to improve the travel efficiency of the whole traffic system, but it leads to the velocity decrease of other vehicles. SG aims to guarantee the personalized driving of each vehicle.

Additionally, the testing results of longitudinal and lateral accelerations are shown in Fig. 14 and Fig. 15, respectively. Aggressive driving style leads to larger acceleration to improve the travel efficiency, and conservative driving style results in larger deceleration to guarantee the driving safety. Larger acceleration and deceleration worsen the ride comfort. The testing results of relative distances are depicted in Fig. 16. Aggressive driving style decreases the safe gap between vehicles, which worsens the driving safety. Instead, conservative driving style shows largest safe gap to advance the driving safety.

\begin{figure}[!t]\centering
	\includegraphics[width=7cm]{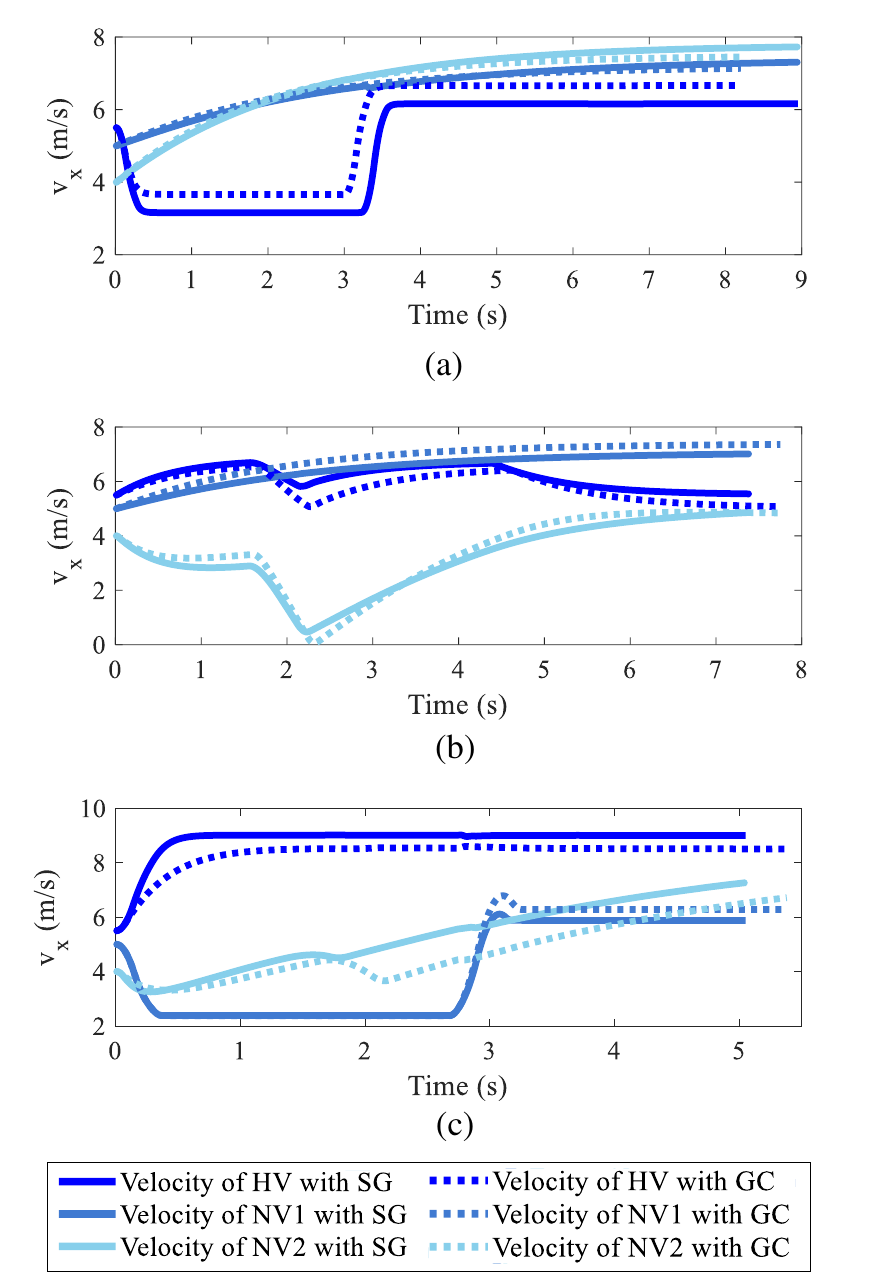}
	\caption{Velocities of CAVs in Case 2: (a) Scenario A; (b) Scenario B; (c) Scenario C.}\label{FIG_12}
\end{figure}

\begin{figure}[!t]\centering
	\includegraphics[width=8.5cm]{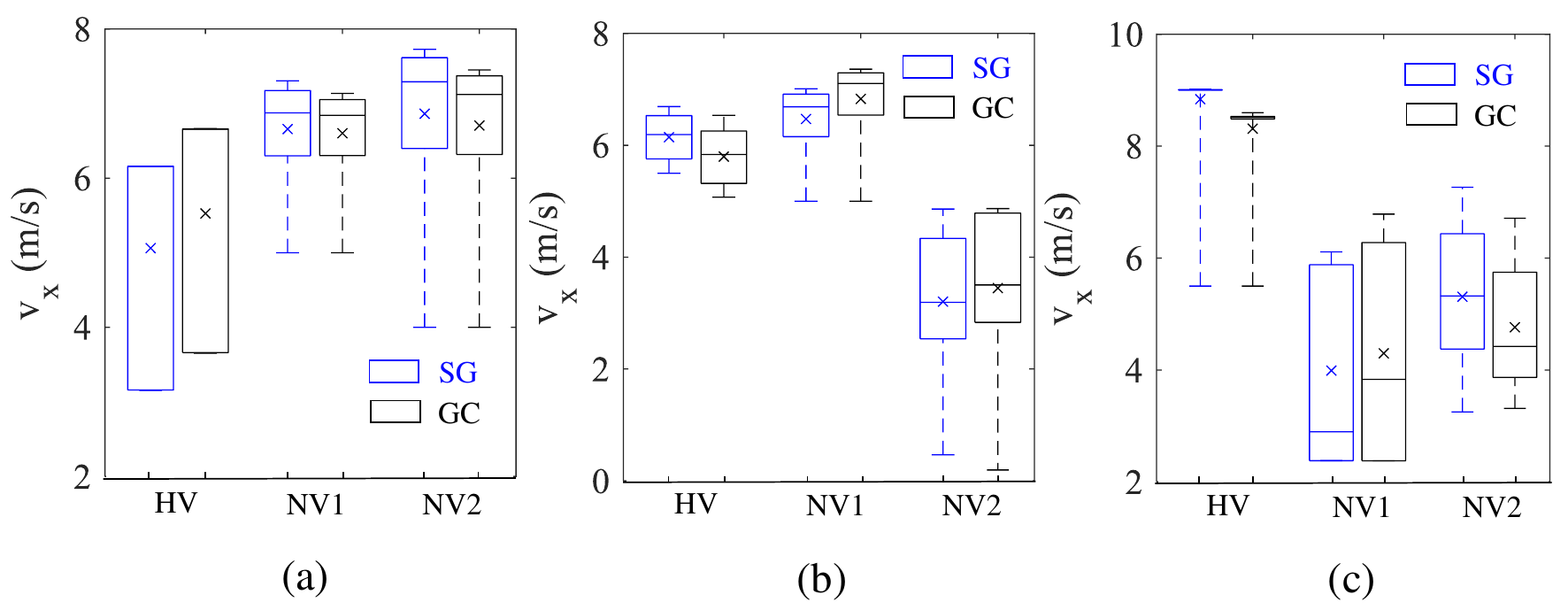}
	\caption{Box plots of velocities in Case 2: (a) Scenario A; (b) Scenario B; (c) Scenario C.}\label{FIG_13}
\end{figure}

\begin{figure}[!t]\centering
	\includegraphics[width=8.5cm]{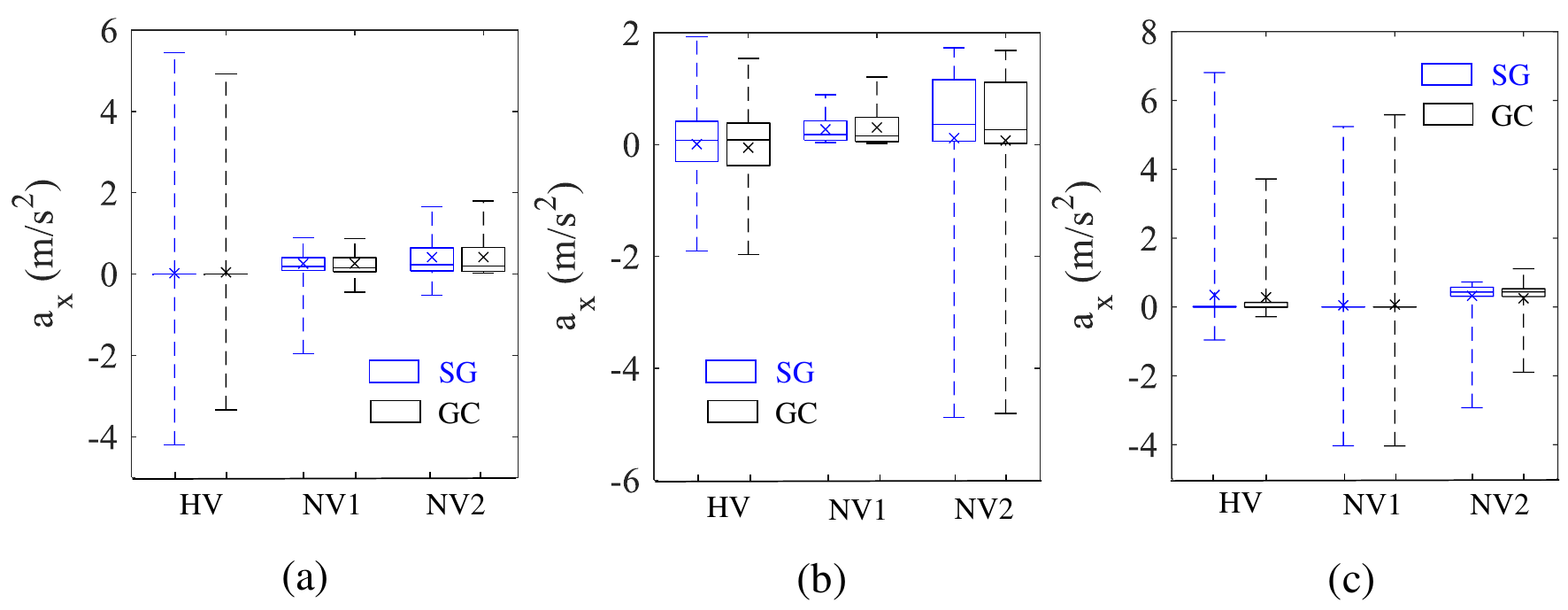}
	\caption{Box plots of longitudinal accelerations in Case 2: (a) Scenario A; (b) Scenario B; (c) Scenario C.}\label{FIG_14}
\end{figure}

\begin{figure}[!t]\centering
	\includegraphics[width=8.5cm]{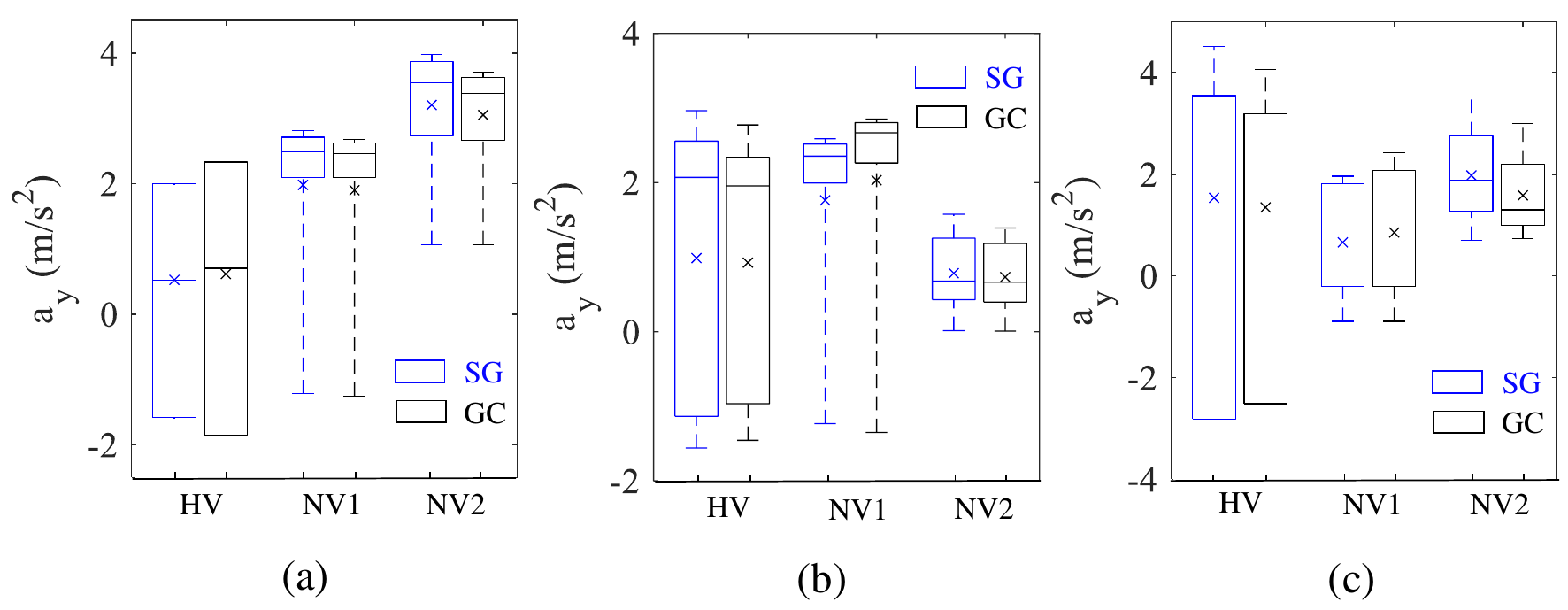}
	\caption{Box plots of lateral accelerations in Case 2: (a) Scenario A; (b) Scenario B; (c) Scenario C.}\label{FIG_15}
\end{figure}

\begin{figure}[!t]\centering
	\includegraphics[width=8.5cm]{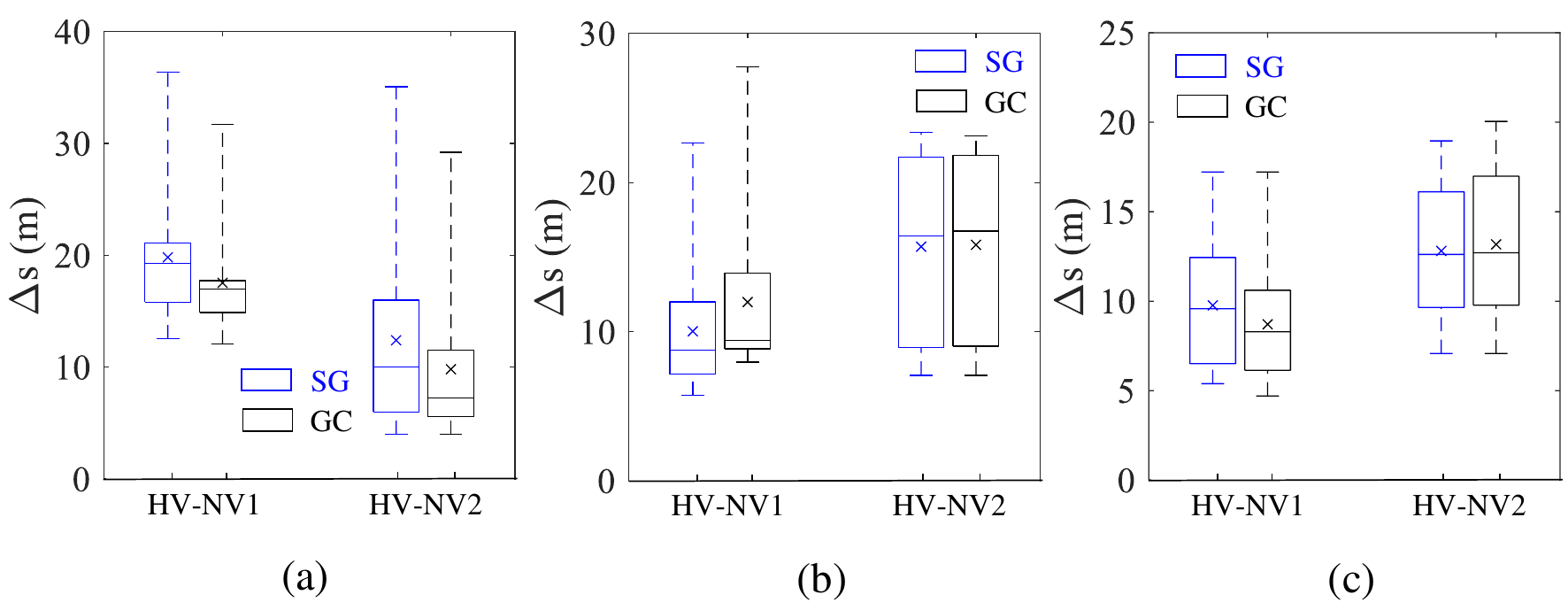}
	\caption{Box plots of relative distances between HV and other NVs in Case 2: (a) Scenario A; (b) Scenario B; (c) Scenario C.}\label{FIG_16}
\end{figure}

\subsection{Testing Case 3}

\begin{figure*}[!t]\centering
	\includegraphics[width=15cm]{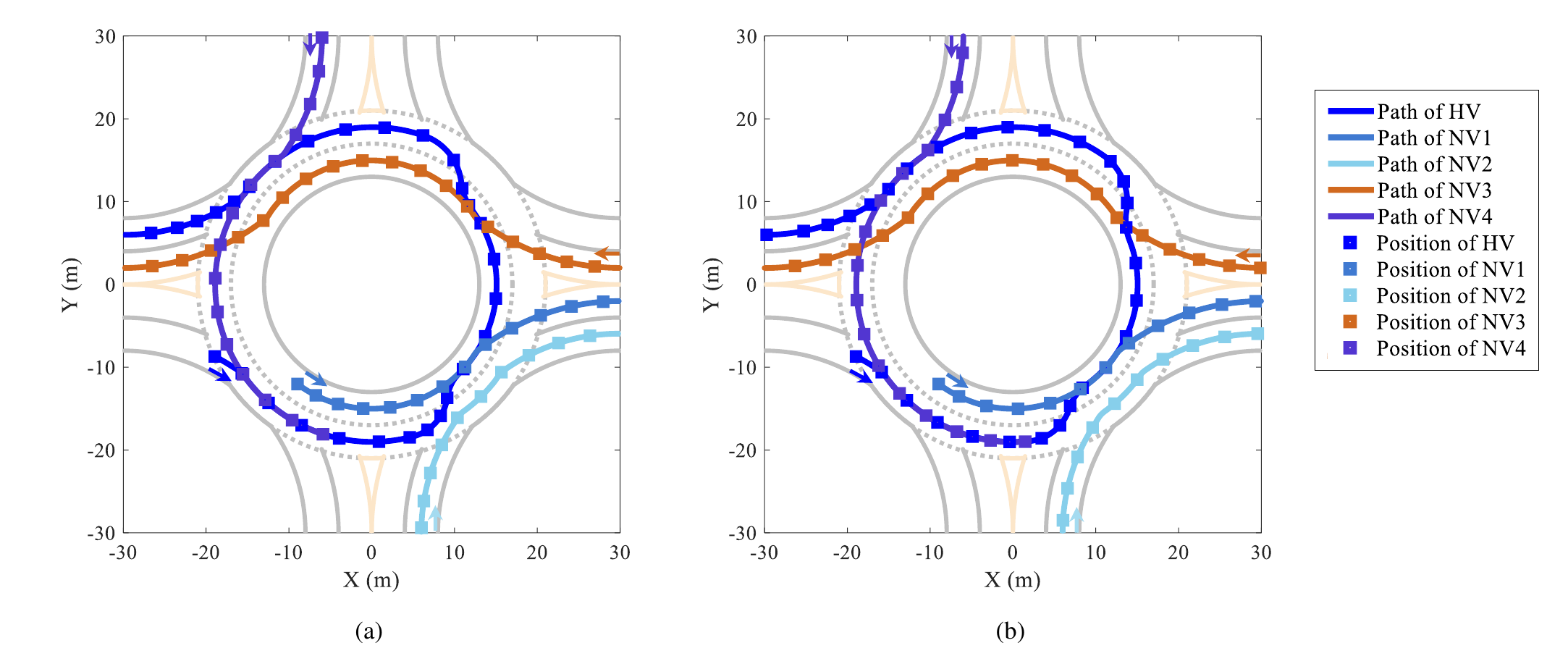}
	\caption{Decision-making results of CAVs in Case 3: (a) SG approach; (b) GC approach.}\label{FIG_17}
\end{figure*}

\begin{table*}[!t]
	\renewcommand{\arraystretch}{1.3}
	\caption{Travel efficiency analysis in Case 3}
\setlength{\tabcolsep}{5 mm}
	\centering
	\label{table_3}
	\resizebox{\textwidth}{10mm}{
		\begin{tabular}{c c c c c c | c c c c c }
			\hline\hline \\[-4mm]
			\multirow{2}{*}{Testing results} & \multicolumn{5}{c|}{SG Approach} & \multicolumn{5}{c}{GC Approach}  \\
\cline{2-11} & \makecell [c] {HV} & \makecell [c] {NV1} & \makecell [c] {NV2} & \makecell [c] {NV3} & \makecell [c] {NV4} & \makecell [c] {HV} & \makecell [c] {NV1} & \makecell [c] {NV2} & \makecell [c] {NV3} & \makecell [c] {NV4}  \\
\hline
			\multicolumn{1}{l}{Velocity Max / (m/s)} & 10.06 & 8.14 & 7.80 & 7.43 & 8.19 & 9.36 & 8.23 & 8.03 & 7.53 & 8.40 \\
			\multicolumn{1}{l}{Velocity RMS / (m/s)} & 8.10 & 7.59 & 7.57 & 6.87 & 7.76 & 7.85 & 8.05 & 7.91 & 7.36 & 8.22 \\
\hline
\multicolumn{1}{l}{System Velocity RMS / (m/s)}  & \multicolumn{5}{c|}{7.59} & \multicolumn{5}{c}{7.88} \\
			\hline\hline
		\end{tabular}
	}
\end{table*}

Case 3 is constructed to evaluate the travel efficiency of the traffic system optimized by the game theoretic decision-making algorithm under multi-vehicle interactions. HV on the outside lane of the main road $M1$ intends to enter the round road $RR$ from the entrance $A_{in}$ and exit from $A_{out}$ to the main road $\hat{M}1$, which is considered as a U-turn scenario. At the roundabout intersection, four NVs are designed. NV1 on the inner lane of the round road $RR$ intends to exit from $C_{out}$ to the main road $\hat{M}3$. NV2 on the external lane of the main road $M2$ wants to enter the round road $RR$ and exit from $C_{out}$ to the main road $\hat{M}3$. NV3 on the inner lane of the main road $M3$ wants to enter the round road $RR$ and exit from $A_{out}$ to the main road $\hat{M}1$. NV4 on the outer lane of the main road $M4$ wants to enter the round road $RR$ from the entrance $D_{in}$ and exit from $D_{out}$ to the main road $\hat{M}4$.

This case also studies the impacts of driving characteristics on the travel efficiency of the traffic system. In this case, the driving style of HV is set as aggressive and the driving styles of four NVs are set as normal. The initial positions of HV, NV1, NV2, NV3 and NV4 are set as (-19, -8.68), (-9, -12), (6, -35), (50, 2) and (-6, 88), respectively. In addition, the initial velocities of HV, NV1, NV2, NV3 and NV4 are set to be 5.5 m/s, 5 m/s, 5 m/s, 4 m/s, 4 m/s, respectively. The performances of the two game theoretic decision making approaches are evaluated and analyzed as well. The testing results are displayed in Figs. 17-19.

It can be found from Fig. 17 that two lane-change decisions are made by HV during its travel at the roundabout to realize collision avoidance and further improve travel efficiency. Fig. 18 shows that, due to the aggressive driving style, frequency acceleration and deceleration behaviors are conducted by HV, which has significant effect on the traffic efficiency. The detailed analysis of travel efficiency is depicted in Table III. Based on the results listed in Table III, system velocity, i.e. the velocity Root Mean Square (RMS) considering all vehicles at the roundabout in real time, is used to evaluate the travel efficiency of the entire traffic system. It can be seen that HV has the largest velocity among all when the SG approach is applied to decision making, which indicates that SG approach can guarantee the personalized driving of CAVs. However, the travel efficiency of the entire traffic system is reduced. The GC approach can improve the system velocity, i.e., the traffic efficiency, but the personalized driving behaviours will be weakened. In real-world practical implementations, the proposed two game theoretic approaches can be used according to different requirements.

Additionally, Fig. 19 shows the computational time of the proposed algorithm in Case 3. The mean value of computational time for each iteration of the SG approach and the GC approach are 0.050s and 0.038s, respectively. For real-time experiments in the future, the computational efficiency can be further improved by using more efficient solver and powerful computing platforms.

\begin{figure}[!t]\centering
	\includegraphics[width=8.8cm]{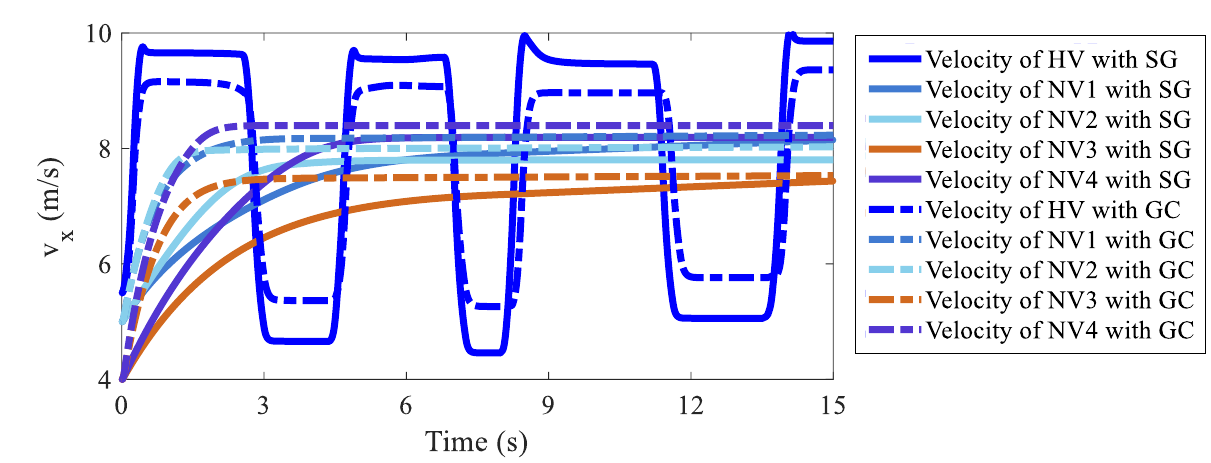}
	\caption{Velocities of CAVs in Case 3.}\label{FIG_18}
\end{figure}

\begin{figure}[!t]\centering
	\includegraphics[width=7.5cm]{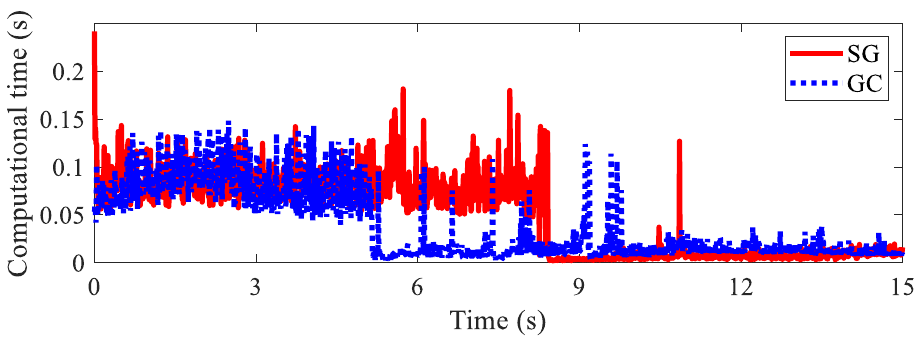}
	\caption{Computational time of the proposed algorithm.}\label{FIG_19}
\end{figure}

\subsection{Discussion}
Based on the analysis of the testing results, it can be concluded that personalized driving characteristic has significant effects on the decision-making results of CAVs. Conservative driving style cares more about driving safety, and aggressive driving style prefers high travel efficiency. The proposed two game theoretic approaches are able to address the decision making issue of CAVs at the unsignalized roundabout. GC tends to improve the traffic efficiency at the expense of weakening the personalized driving. SG focuses on guaranteeing the personalized driving without considering the system efficiency. In general, the two game theoretic approaches can be applied to different scenarios according different demands. In different traffic scenarios, the decision-making cost functions and behavior representation variables could be different, but the designed high-level game theoretic decision framework is still applicable. The lane change and merging scenarios on highway have been tested with the proposed decision-making framework in the previous works [39,40]. The proposed decision-making algorithm will be further improved in the future to address more complex traffic scenarios.

Besides, with the increase of the number of players in a game, the computation burden would increase. This issue can be addressed in the future by improving the computation efficiency of the solver and using high-performance computing resources.

\section{Conclusion}
In this paper, two game theoretic approaches are proposed to deal with the decision making issue of CAVs at the unsignalized roundabout considering personalized driving. In the decision-making framework, the motion prediction module is designed to advance the algorithm performance. In the definition of the decision-making payoff function, different driving styles of CAVs are considered, which are related to the performances of safety, comfort and efficiency. Furthermore, Stackelberg game and grand coalition game are applied to the decision making of CAVs at the unsignalized roundabout. To verify the proposed decision-making algorithm, two testing cases are designed and conducted considering personalized driving of CAVs. According to the testing results, it is found that the two game theoretic approaches can make feasible and reasonable decisions for CAVs at the unsignalized roundabout, and personalized driving leads to different decision-making results.
The testing results indicate that the proposed algorithm is capable to ensure the safety of the traffic system at the complex roundabout zone, and meanwhile guarantee the personalized driving demands of individuals within the area.

Our future work will focus on the improvement of the proposed algorithm considering more complex traffic scenarios including the interactions between CAV and human-driven vehicles, and the algorithms will be further applied in real-time vehicle onboard hardware systems for validation and refinement.

\end{document}